\documentclass{article}

\usepackage[english]{babel}

\usepackage[letterpaper,top=2cm,bottom=2cm,left=3cm,right=3cm,marginparwidth=1.75cm]{geometry}

\usepackage{amsmath}
\usepackage{amssymb} 
\usepackage{graphicx}
\usepackage[colorlinks=true, allcolors=blue]{hyperref}
\usepackage{rotating}
\usepackage{placeins}
\usepackage{physics}
\usepackage{siunitx}
\usepackage{float}  %
\usepackage{authblk}
\usepackage{xcolor}
\usepackage[numbers]{natbib}
\usepackage{hyperref}
\usepackage{soul}
\usepackage{bibunits}
\usepackage{bm}
\usepackage[T1]{fontenc}

\hypersetup{
    colorlinks=true,
    linkcolor=red,      
    citecolor=blue,      
    urlcolor=blue       
}

\hypersetup{linktocpage,colorlinks=true,citecolor=blue}

 \DeclareSIUnit\mt{\milli\tesla}
\title{Field-controlled breaking and restoration of parity-time symmetry in Josephson interference} 

\author[1,2,3]{Yi-Chen Tsai}
\author[1]{Yung-Yeh Chang}
\author[1,4]{Tao-Yi Hsu}
\author[1,4]{Thomas Kuo}
\author[5,6,7]{Chia-Nung Kuo}
\author[5,6,7]{Chin-Shan Lue}
\author[4,8]{Kuei-Lin Chiu}
\author[1,9,*]{Chen-Hsuan Hsu}
\author[1,10,*]{Chung-Ting Ke}

\affil[1]{Institute of Physics, Academia Sinica, Taipei 115201, Taiwan}
\affil[2]{Department of Engineering and System Science, National Tsing Hua University, Hsinchu 300044, Taiwan}
\affil[3]{Nanoscience and Technology Program, Taiwan International Graduate Program, Academia Sinica, Taipei 115201, Taiwan}
\affil[4]{Department of Physics, National Sun Yat-Sen University, Kaohsiung 804, Taiwan}
\affil[5]{Program on Key Materials, Academy of Innovative Semiconductor and Sustainable Manufacturing (AISSM), National Cheng Kung University, Tainan 70101, Taiwan}
\affil[6]{Department of Physics, National Cheng Kung University, Tainan 701, Taiwan}
\affil[7]{Taiwan Consortium of Emergent Crystalline Materials (TCECM), National Science and Technology Council, Taipei 10601, Taiwan}
\affil[8]{Research Center for Quantum Computing and Quantum Materials, National Sun Yat-Sen University, Kaohsiung 80424, Taiwan}
\affil[9]{Physics Division, National Center for Theoretical Sciences, Taipei 106319, Taiwan}
\affil[10]{Research Center for Critical Issues, Tainan 711010, Taiwan}
\affil[*]{Corresponding authors: chenhsuan@as.edu.tw, ctke@as.edu.tw}

\begin{document}
\begin{bibunit}[unsrt]
\maketitle

\begin{abstract}

Symmetry plays a fundamental role in determining the phases and physical properties of quantum matter. Controlling symmetry in mesoscopic superconducting devices provides a route to reconfigure their phase-coherent transport.
Here we demonstrate symmetry-selective Josephson interferometry in lateral NbTi/PtTe$_{\text{2}}$/NbTi junctions by controlling the relative orientations of the current and magnetic field. From the supercurrent interference patterns, we construct a field-current symmetry map that identifies 
configurations exhibiting or violating the device-level parity
($\mathcal{P}$), time-reversal ($\mathcal{T}$) and their combined $\mathcal{PT}$ symmetry. In the absence of an in-plane field, the junction exhibits a symmetric Fraunhofer pattern. An in-plane field parallel to the current produces a pronounced side-lobe asymmetry, whereas reversing both the current and the complete magnetic-field configuration restores a generalized $\mathcal{T}$ relation. Remarkably, orienting the in-plane field perpendicular to the current restores the $\mathcal{PT}$-symmetric Fraunhofer response even at substantial field strengths. A microscopic model attributes this behavior to the interplay between disorder-induced potential variations and flux dipoles generated by in-plane-field Meissner focusing near the superconducting electrodes. Our results establish a reconfigurable Josephson interferometer in which the field-current geometry selects the symmetry operation being probed and switches the device between symmetry-broken and symmetry-restored interference states.
 
\end{abstract}

\section*{Introduction} 

Symmetry provides a unifying framework for understanding quantum systems, from the gauge principles underlying particle interactions to the classification of phases in condensed matter. Breaking or continuously tuning a symmetry can enable phases and transport responses that are forbidden in more symmetric configurations. Familiar examples include time-reversal ($\mathcal{T}$) symmetry breaking in ferromagnets~\cite{Wang2026} and $U(1)$ symmetry breaking in superconductors~\cite{BCS1957}. 
These phenomena illustrate how symmetry control can be used both to engineer quantum states and to uncover unconventional  responses~\cite{Wang2026,Trahms2023,21vm-n8gp}.

Quantum devices provide a natural setting in which such control can be implemented and probed directly through transport. Superconductor-normal-superconductor (SNS) Josephson junctions (JJs) are particularly versatile because their supercurrent and current-phase relation are shaped by the properties of the normal region and by external electric and magnetic fields~\cite{DynesFulton:1971,GolubovRMP:2004,AmundsenRMP:2024}.
This interplay enables the manipulation of parity ($\mathcal{P}$) and $\mathcal{T}$ symmetries across a range of material platforms, 
including electrostatically defined and symmetry-broken JJs in magic-angle twisted bilayer graphene~\cite{deVriesNatNano:2021,diez2023symmetry}, unconventional supercurrent phases in NbSe$_2$-based Ising-superconductor heterostructures~\cite{IdzuchiNatComm:2021}, spin-orbit- and field-controlled phase shifts and interference patterns in planar JJs~\cite{Flensberg:2016PRB,MayerNatComm:2020,ReinhardtNatComm:2024}, and gate- and flux-engineered higher-harmonic current-phase relations in planar Ge junctions~\cite{LeblancNatComm:2025}.
The Josephson diode effect further demonstrates how broken symmetries can generate nonreciprocal superconducting transport~\cite{NadeemNatRevPhys:2023,Baumgartner2022NatNano,Chirolli2025DisorderedFraunhofer,Parkin:2022JDE}. Yet a systematic means of selectively probing, breaking and restoring $\mathcal{P}$ and $\mathcal{T}$ symmetries within a single JJ remains lacking. Such control would establish symmetry itself as a reconfigurable degree of freedom in supercurrent transport.

In this work, we demonstrate symmetry-selective control of Josephson interference in a PtTe$_2$ JJ using the magnetic-field orientation and current polarity. Under a purely out-of-plane field, the junction exhibits a conventional symmetric Fraunhofer pattern, consistent with preserved $\mathcal{P}$ and $\mathcal{T}$ symmetries and an approximately uniform supercurrent-density distribution. Introducing an in-plane field produces a pronounced asymmetric interference pattern already below \SI{10}{\milli\tesla}. A generalized $\mathcal{T}$ relation is recovered upon simultaneously reversing the current and the complete magnetic-field configuration, demonstrating that the interference symmetry is governed by the relative field-current orientation. This response persists up to approximately \SI{100}{\milli\tesla} and exhibits a pronounced two-fold angular dependence. In particular, orienting the in-plane field perpendicular to the current restores the $\mathcal{PT}$-symmetric Fraunhofer response despite finite $B_{\rm in}$, defining $\mathcal{PT}$-invariant boundaries in the field-current symmetry map.

To identify the microscopic mechanism, we develop an effective model of supercurrent transport in the PtTe$_2$ JJ. The asymmetric pattern arises from the interplay between in-plane-field flux focusing and disorder in the diffusive transport channel. Owing to the Meissner effect, the in-plane field is distorted near the superconducting electrodes, generating additional out-of-plane flux components of opposite signs at the two normal-superconductor (NS) interfaces. These local flux dipoles modulate the phase accumulated along the supercurrent trajectories. In the presence of disorder-induced spatial variations in the local potential, reversing the out-of-plane field $B_{ z}$ changes the dipole-induced flux sampled by individual trajectories, producing an asymmetric interference pattern. The dipole contribution is strongest when the in-plane field is parallel or antiparallel to the current ($\theta=0^\circ$ and $180^\circ$) and is suppressed when the field is perpendicular ($\theta=90^\circ$ and $270^\circ$). The model reproduces the principal experimental trends and establishes a controllable mechanism for engineering the symmetry of Josephson interference.

\section*{Field-current control of Josephson interference symmetry}
 
\begin{figure}[!tp]
\centering
\includegraphics[width=1\linewidth]{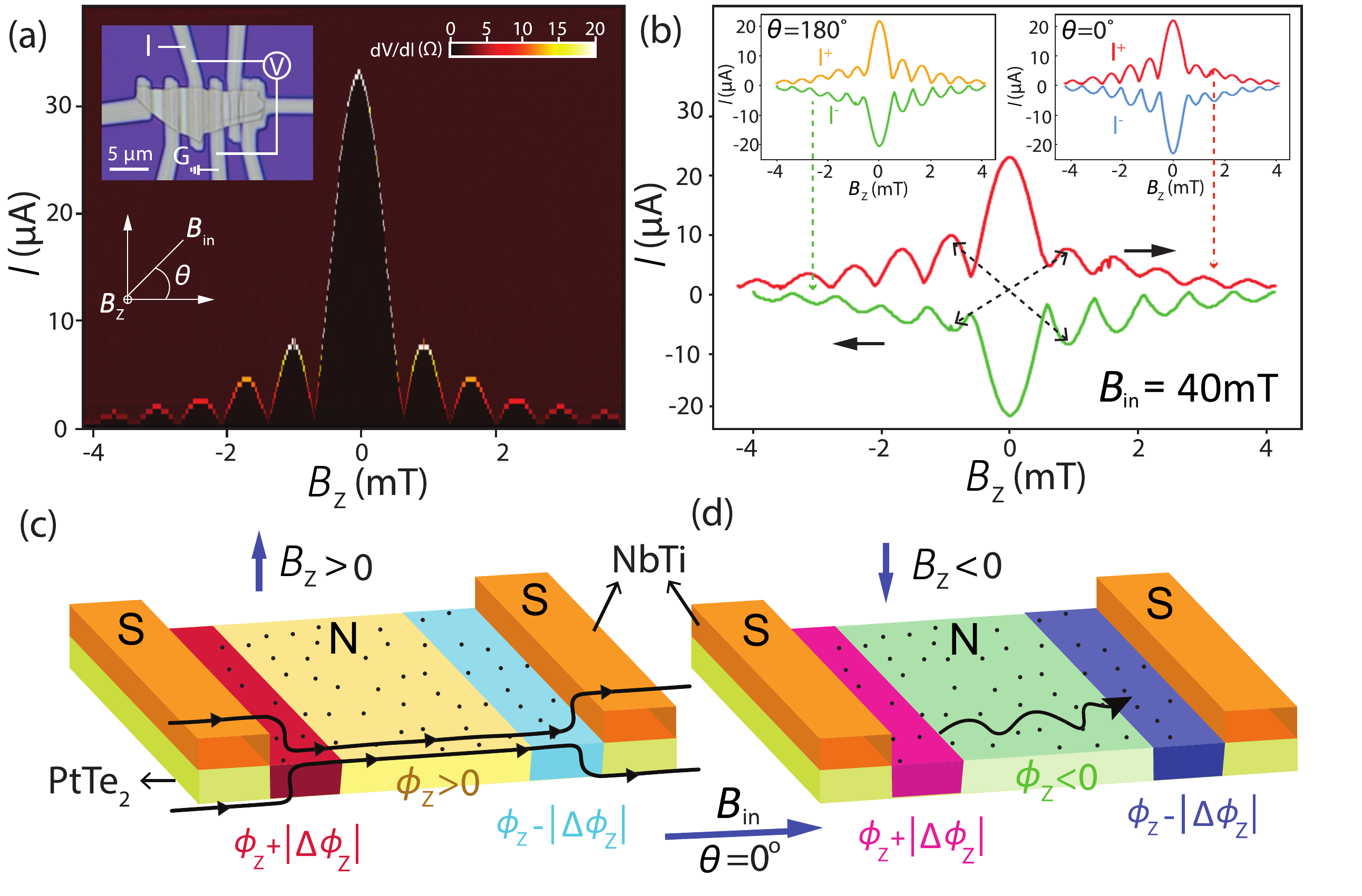} 
\caption{
\textbf{Measurement setup of Josephson interference  and microscopic origin of the asymmetry.}
(a)  Differential resistance $dV/dI$ as a function of $B_{z}$ and bias current, measured in the absence of an in-plane field. The symmetric Fraunhofer interference pattern, with well-defined central and side lobes, is consistent with an approximately uniform supercurrent distribution across the junction.
Inset: optical image of the NbTi/PtTe$_2$/NbTi JJs with the measurement configuration overlaid. 
(b) Interference patterns measured at $B_{\text{in}}=\SI{40}{\milli\tesla}$ for $I^+$ at $\theta=0^\circ$ and $I^-$ at $\theta=180^\circ$, where $I^{\pm}$ denote the positive and negative current directions, respectively. Upon reversal of $B_z$, the two profiles map onto one another, illustrating the generalized $\mathcal{T}$ relation under simultaneous reversal of the current and in-plane-field direction. The left (right) inset shows the corresponding interference pattern at $\theta=180^\circ$ ($\theta=0^\circ$), also measured at $B_{\text{in}}=\SI{40}{\milli\tesla}$.
(c,d) Microscopic origin of the asymmetric interference pattern arising from the interplay between disorder and flux dipoles. The yellow (green) region denotes the normal region subject to a uniform background out-of-plane flux $\phi_z>0$ ($\phi_z<0$) away from the superconducting electrodes. Owing to flux focusing, the in-plane field lines bend near the two NS interfaces [black curves in (c)], generating finite out-of-plane flux contributions of opposite signs, as indicated by the red and blue stripes. The black  curve in (d) represents a carrier trajectory. 
In the presence of disorder, reversing the sign of $\phi_z$ changes the dipole-induced flux sampled along the trajectory, resulting in different accumulated phases and hence an asymmetric interference pattern. 
}

\label{fig:Fig1}
\end{figure}

Our JJ devices are fabricated from PtTe$_{2}$  in the 1T phase. Although 1T-PtTe$_{2}$ is a centrosymmetric type-II Dirac semimetal~\cite{MingzheNatCommPtTe2:2017,Dongzhi2018:PtTe-QuantumOscil,Zheng2018:PtTe2-FS}, the Fermi level in our devices lies far from the Dirac point, resulting in a highly doped transport channel with semiconducting behavior. PtTe$_{2}$ flakes with thicknesses of \SI{20}{\nano\meter}--\SI{40}{\nano\meter} are contacted by sputtered NbTi electrodes patterned using conventional electron-beam lithography to form lateral SNS JJs, as shown in the inset of Fig.~\ref{fig:Fig1}(a). Full device geometries and fabrication details are provided in the Methods and  Supplementary Information (SI) Sec.~\ref{sec:Device}. 
The devices are measured in a dilution refrigerator with a base temperature of approximately \SI{50}{\milli\kelvin}, equipped with a three-axis vector magnet. We define the magnetic-field configuration by an out-of-plane component $B_{z}$ and an in-plane field of magnitude $B_{\text{in}}$ oriented at an angle $\theta$ relative to the junction direction. Here, $\theta=0^\circ$ corresponds to the junction-current direction, as indicated in the inset of Fig. \ref{fig:Fig1}(a).

In the absence of an in-plane field, the supercurrent exhibits a conventional symmetric Fraunhofer interference pattern, with a central lobe spanning approximately two flux quanta and evenly spaced side lobes with decaying amplitudes, as shown in Fig.~\ref{fig:Fig1}(a). The symmetry of this pattern is consistent with an approximately uniform supercurrent-density distribution and establishes the baseline response of the junction. At the device level, the interference pattern directly probes the $\mathcal{P}$ and $\mathcal{T}$  relations~\cite{KTLawPRB:2018AJE,Kashiwaya:2019PRB}:
\begin{eqnarray}
\mathcal{P}: && I^{\pm}(B_{z})=-I^{\mp}(B_{z}),
\label{Eq:P-symmetry_main}
\\
\mathcal{T}: && I^{\pm}(B_{z})=-I^{\mp}(-B_{z}),
\label{Eq:T-symmetry_main}
\end{eqnarray}
where $I^{\pm}$ denote the positive and negative current branches, respectively. Combining these relations, $\mathcal{PT}$ symmetry implies 
$
I^{\pm}(B_{z})=I^{\pm}(-B_{z}),
$
corresponding to a symmetric Fraunhofer pattern under reversal of the out-of-plane field.
Here, $\mathcal{P}$ and $\mathcal{T}$ denote operational symmetries of the measured device response, rather than exact symmetries of the microscopic Hamiltonian (see SI Sec.~\ref{Sec:microscopic}).

Applying an in-plane field along the current direction produces a pronounced asymmetric interference pattern, as shown for $B_{\text{in}}=\SI{40}{\milli\tesla}$ in Fig.~\ref{fig:Fig1}(b). In addition to an overall suppression of the critical current, most prominently in the central lobe, the side lobes develop a clear imbalance in amplitude. This asymmetry persists across the measured side lobes, while the amplitudes on one side retain a decay similar to that of a conventional Fraunhofer pattern. Reversing the in-plane-field direction produces a mirror-reflected lobe asymmetry, as shown in the insets of Fig.~\ref{fig:Fig1}(b). By comparing the interference patterns measured for opposite current directions at fixed $B_{\text{in}}$ and $\theta$, we find that the relation in Eq.~\eqref{Eq:P-symmetry_main}  remains fulfilled.
 
A distinct correspondence emerges when the current and in-plane field are reversed simultaneously. As shown in Fig.~\ref{fig:Fig1}(b), the interference profile measured for $I^{+}$ with $B_{\text{in}}$ at $\theta=0^\circ$ (red) maps onto that measured for $I^{-}$ at $\theta=180^\circ$ (green) upon reversing $B_z$. Accordingly, the suppressed right side lobes in the positive-current branch correspond to the suppressed left side lobes in the negative-current branch. This diagonal symmetry shows that the interference pattern is governed not only by the relative orientation of the perpendicular field and current, but also by that of the in-plane field, and obeys
\begin{equation}
I^{\pm}  (B_{z}, B_{\text{in}}, \theta) = -I^{\mp} 
(-B_{z}, B_{\text{in}}, \theta + 180^\circ).
\label{Eq:general-T-symmetry}
\end{equation}
The same correspondence is observed in two junctions with perpendicular current directions fabricated on the same PtTe$_2$ flake, indicating that the response does not originate from an intrinsic crystallographic anisotropy (SI Sec.~\ref{sec:additional_data}). 

To account for the observed asymmetry, we employ a microscopic model incorporating two essential ingredients~\cite{FlensbergPRB:2017_AnomalousFraunhoferPattern}: flux dipoles generated by the in-plane magnetic field and disorder-induced local potential variations that give rise to diffusive transport, as illustrated in Figs.~\ref{fig:Fig1}(c) and (d). Owing to the Meissner effect and the geometry of the approximately \SI{70}{\nano\meter}-thick superconducting electrodes, the in-plane field lines are distorted near the contacts, generating additional out-of-plane flux components of opposite signs at the two NS interfaces [Fig.~\ref{fig:Fig1}(c)]. These flux dipoles contribute an additional phase $\Delta\phi_{z}$ to the phase $\phi_{z}$ induced by $B_{z}$. For $B_{\text{in}}$ applied at $\theta=0^\circ$, the total phases in the red and blue regions are $\phi_{z}+|\Delta\phi_{z}|$ and $\phi_{z}-|\Delta\phi_{z}|$, respectively, as illustrated for negative $B_{z}$ in Fig.~\ref{fig:Fig1}(d). Disorder-induced potential variations, represented by the dots in the channel, generate spatially nonuniform supercurrent trajectories that sample the two dipole regions differently. Reversing $B_{z}$ therefore changes the dipole-induced flux sampled along these trajectories, resulting in different accumulated phases and hence an asymmetric interference pattern. The flux-dipole contribution is strongly suppressed when the in-plane field is perpendicular to the current.

Guided by this microscopic picture, we next probe the $\mathcal{P}$ and $\mathcal{T}$ symmetries, as well as their combined $\mathcal{PT}$ symmetry, by varying the orientation of the in-plane field.

\section*{Symmetry-selective control through in-plane-field rotation} 
 
\begin{figure}[!t]
\centering
\includegraphics[width=1\linewidth]{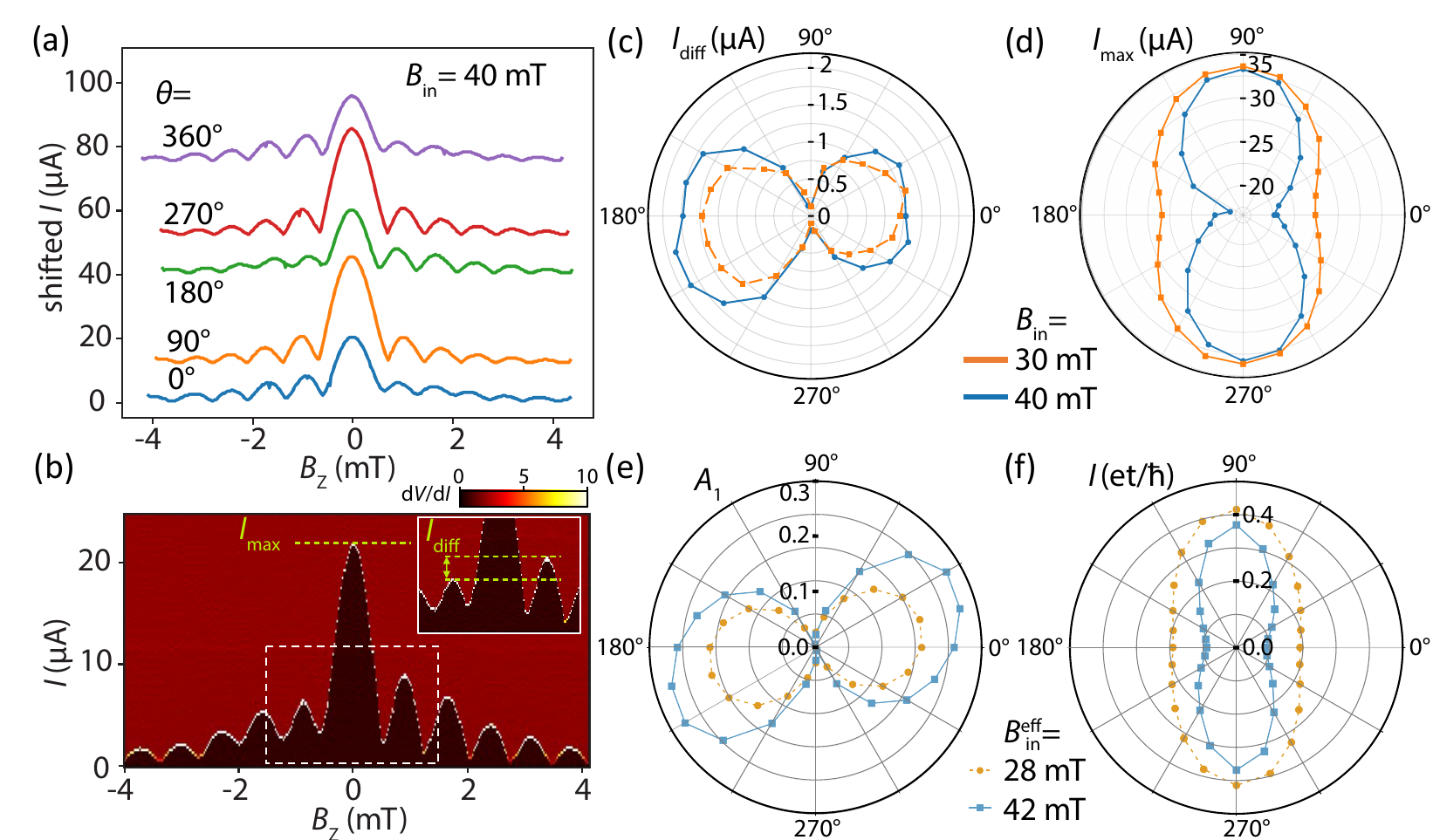} 
\caption{\textbf{Symmetry-selective analysis of Josephson interference by in-plane-field rotation.} 
(a) Critical-current interference patterns measured at a fixed in-plane field magnitude $B_{\rm in}=\SI{40}{\milli\tesla}$ while rotating its orientation $\theta$. The side-lobe asymmetry reverses between $\theta=0^\circ$ and $180^\circ$, whereas the patterns become nearly symmetric at $\theta=90^\circ$ and $270^\circ$.
(b) Representative $dV/dI$ map. Inset: extraction of the first-side-lobe difference $I_{\mathrm{diff}}$ and the central-lobe maximum $I_{\mathrm{max}}$. 
(c,d) Angular dependences of the experimentally extracted quantities for $B_{\rm in}=\SI{30}{\milli\tesla}$ and \SI{40}{\milli\tesla}. (c) The magnitude $|I_{\mathrm{diff}}|$ exhibits a two-fold angular dependence and is largest when the in-plane field is parallel or antiparallel to the current. (d) The central-lobe maximum $I_{\mathrm{max}}$ exhibits an anticorrelated two-fold response and is largest when the field is perpendicular to the current.
(e,f) Calculated angular dependences of  (e) the first-lobe asymmetry $A_1$ and (f) the central-lobe critical current for two effective in-plane field strengths, $B_{\rm in}^{\rm eff}\approx\SI{28}{\milli\tesla}$ and \SI{42}{\milli\tesla}. The critical current in panel (f) is shown in units of $et/\hbar$, where $t$ is the nearest-neighbor hopping strength in the normal region. The effective fields account for the difference between the simulated and experimental system sizes (see Methods). 
}
\label{fig:Fig2}
\end{figure}

To resolve the directional dependence of the interference pattern, we fix the current direction and $B_{\text{in}}$ while rotating the in-plane-field orientation $\theta$. Figure~\ref{fig:Fig2}(a) shows representative patterns at $\theta=0^\circ$, $90^\circ$, $180^\circ$, and $270^\circ$. The asymmetry is strongest when the field is parallel or antiparallel to the current, accompanied by a pronounced suppression of the central-lobe maximum $I_{\text{max}}$. Rotating the field from $\theta=0^\circ$ to $180^\circ$ reverses the lobe asymmetry, whereas nearly symmetric patterns with restored $I_{\text{max}}$ emerge at $\theta=90^\circ$ and $270^\circ$. Remarkably, these field-current configurations also define a symmetry-selective measurement protocol. At fixed $B_{\text{in}}$ and $\theta$, comparison of the opposite current branches probes $\mathcal{P}$ symmetry, as illustrated in the insets of Fig.~\ref{fig:Fig1}(b). By contrast, the generalized $\mathcal{T}$ symmetry is tested by simultaneously reversing the current and the complete magnetic-field configuration, $B_{z}\rightarrow-B_{z}$ and $\theta\rightarrow\theta+180^\circ$, according to Eq.~\eqref{Eq:general-T-symmetry}. The field-current geometry therefore determines not only the interference profile, but also the symmetry operation being interrogated.

We next perform this symmetry-selective analysis by quantifying the continuous angular evolution of the interference pattern.
The side-lobe asymmetry is characterized by $I_{\mathrm{diff}}$, defined as the difference between the peak supercurrents of the first side lobes at positive and negative $B_{z}$, as indicated by the dashed boxes in Fig.~\ref{fig:Fig2}(b). To reduce sensitivity to stochastic switching events, each peak value is obtained by averaging the 20 data points near the maximum of the corresponding lobe; see the inset of Fig.~\ref{fig:Fig2}(b). We also extract the central-lobe maximum $I_{\mathrm{max}}$, indicated in the same panel. Figures~\ref{fig:Fig2}(c) and \ref{fig:Fig2}(d) show the angular dependences of $I_{\mathrm{diff}}$ and $I_{\mathrm{max}}$ for $B_{\text{in}}=\SI{30}{\milli\tesla}$ and \SI{40}{\milli\tesla}. Both quantities exhibit a clear two-fold angular periodicity, but with anticorrelated responses: $I_{\mathrm{diff}}$ reaches its extrema and $I_{\mathrm{max}}$ is suppressed when the in-plane field is parallel or antiparallel to the current ($\theta=0^\circ$ or $180^\circ$), whereas the asymmetry is minimized and $I_{\mathrm{max}}$ is largest when the field is perpendicular ($\theta=90^\circ$ or $270^\circ$).

These angular trends support the microscopic picture in which the asymmetry originates from dipole-induced phase accumulation. The stronger modulation observed at $B_{\rm in}=\SI{40}{\milli\tesla}$ than at $B_{\rm in}=\SI{30}{\milli\tesla}$ shows that the effect increases with the in-plane field, as expected for a flux-dipole mechanism. Similar angular dependences are observed in devices with different junction lengths and current directions relative to the PtTe$_{\rm 2}$ crystal axes, indicating that the response is governed by the field-current geometry rather than by sample-specific or crystallographic anisotropy (SI Sec.~\ref{sec:additional_data}). 

To test this microscopic interpretation, we calculate the supercurrent and extract the first-lobe asymmetry and the central-lobe critical current, shown in Figs.~\ref{fig:Fig2}(e) and \ref{fig:Fig2}(f), respectively. We define the asymmetry of the $n$-th lobe as
$
A_n \equiv \left| \left[ I^{(n)} -I^{(-n)} \right] / \left[ I^{(n)}+I^{(-n)} \right] \right|
$
where $I^{(\pm n)}$ denote the critical currents of the $n$-th lobes at positive and negative $B_{z}$, respectively. The calculations reproduce the experimentally observed angular dependences of both quantities [Figs.~\ref{fig:Fig2}(c)--\ref{fig:Fig2}(f)].  For parameters relevant to the experiment, the principal axis of maximal asymmetry remains close to the junction direction at $\theta=0^\circ$ and $180^\circ$.

\begin{figure}[htbp]
\centering
\includegraphics[width=1\linewidth]{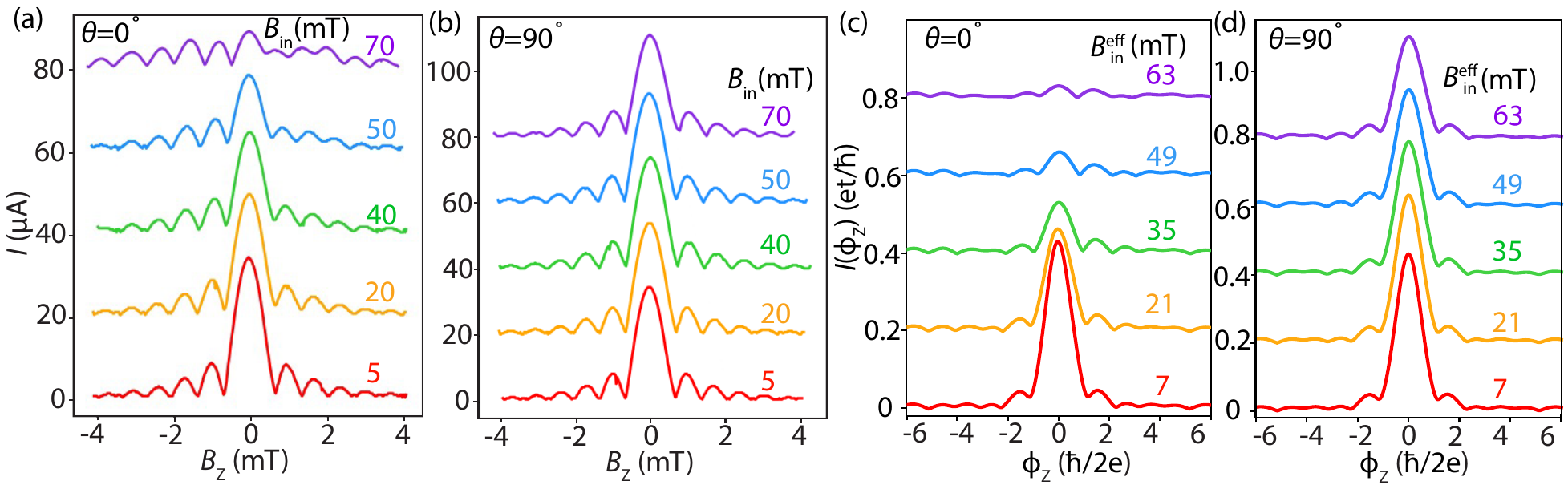}
\caption{\textbf{Orientation-selective restoration of symmetric Josephson interference.} 
(a,b) Interference patterns measured at several in-plane field strengths. (a) For an in-plane field parallel to the current ($\theta=0^\circ$), the side-lobe asymmetry increases with $B_{\rm in}$, accompanied by suppression of the central lobe and an eventual evolution towards a SQUID-like pattern. (b) For an in-plane field perpendicular to the current ($\theta=90^\circ$), the interference pattern remains essentially symmetric over the measured field range, demonstrating the robust restoration of the $\mathcal{PT}$-symmetric Fraunhofer response despite a finite $B_{\rm in}$.
(c,d) Calculated interference patterns $I(\phi_z)$ as a function of the perpendicular magnetic flux $\phi_z$, expressed in units of the flux quantum, for (c) $\theta=0^\circ$ and (d) $\theta=90^\circ$. The calculations reproduce the contrasting field evolution observed experimentally. The corresponding effective in-plane field strengths $B_{\rm in}^{\rm eff}$ are indicated and account for the difference between the simulated and experimental system sizes (see Methods and SI Sec.~\ref{Sec:microscopic}). 
}
\label{fig:Fig3}
\end{figure}

The agreement between experiment and theory establishes in-plane-field rotation as a symmetry-selective protocol for probing the generalized $\mathcal{T}$ symmetry. It also identifies $\theta=90^\circ$ and $270^\circ$ as configurations in which the $\mathcal{PT}$-symmetric Fraunhofer response is restored.
We next examine the robustness of this restoration against the in-plane field strength by comparing fields oriented parallel and perpendicular to the current [Figs.~\ref{fig:Fig3}(a) and \ref{fig:Fig3}(b)]. For the parallel configuration, the lobe asymmetry increases progressively with $B_{\rm in}$, accompanied by a marked suppression of the central lobe and an eventual evolution towards a SQUID-like pattern (see Fig.~\ref{Fig:evolution} in SI Sec.~\ref{sec:additional_data}). Similar Fraunhofer-to-SQUID-like crossovers have previously been discussed in connection with field-induced finite-momentum pairing and Zeeman-driven $0$--$\pi$ transitions in spin-orbit-coupled JJs~\cite{HartNatPhys:2017,ChenNatComm:2018,DvirPRB:2021,PalNatPhys:2022}. In our devices, however, the observed angular dependence and its agreement with the microscopic model instead support a mechanism governed predominantly by flux focusing and disorder.

By contrast, for the perpendicular configuration, the interference pattern remains essentially symmetric up to $B_{\rm in}=\SI{70}{\milli\tesla}$ [Fig.~\ref{fig:Fig3}(b)]. Thus, selecting $\theta=90^\circ$ or $270^\circ$ restores the $\mathcal{PT}$-symmetric Fraunhofer response characteristic of $B_{\rm in}=0$, despite the presence of a substantial in-plane field. This orientation-selective restoration is consistent with the suppression of the flux-dipole-induced phase modulation when the in-plane field is perpendicular to the current, demonstrating robust symmetry control over a broad field range.

The theoretical simulations reproduce the contrasting field-dependent responses for both orientations, as shown in Figs.~\ref{fig:Fig3}(c) and \ref{fig:Fig3}(d). For $\theta=0^\circ$, increasing $B_{\rm in}$ suppresses the central lobe and drives the interference pattern towards a SQUID-like form, whereas for $\theta=90^\circ$, the symmetric Fraunhofer pattern remains comparatively stable. This agreement further supports the orientation-selective restoration of the $\mathcal{PT}$-symmetric interference response.

\begin{figure}[!t]
\centering
\includegraphics[width=1\textwidth]{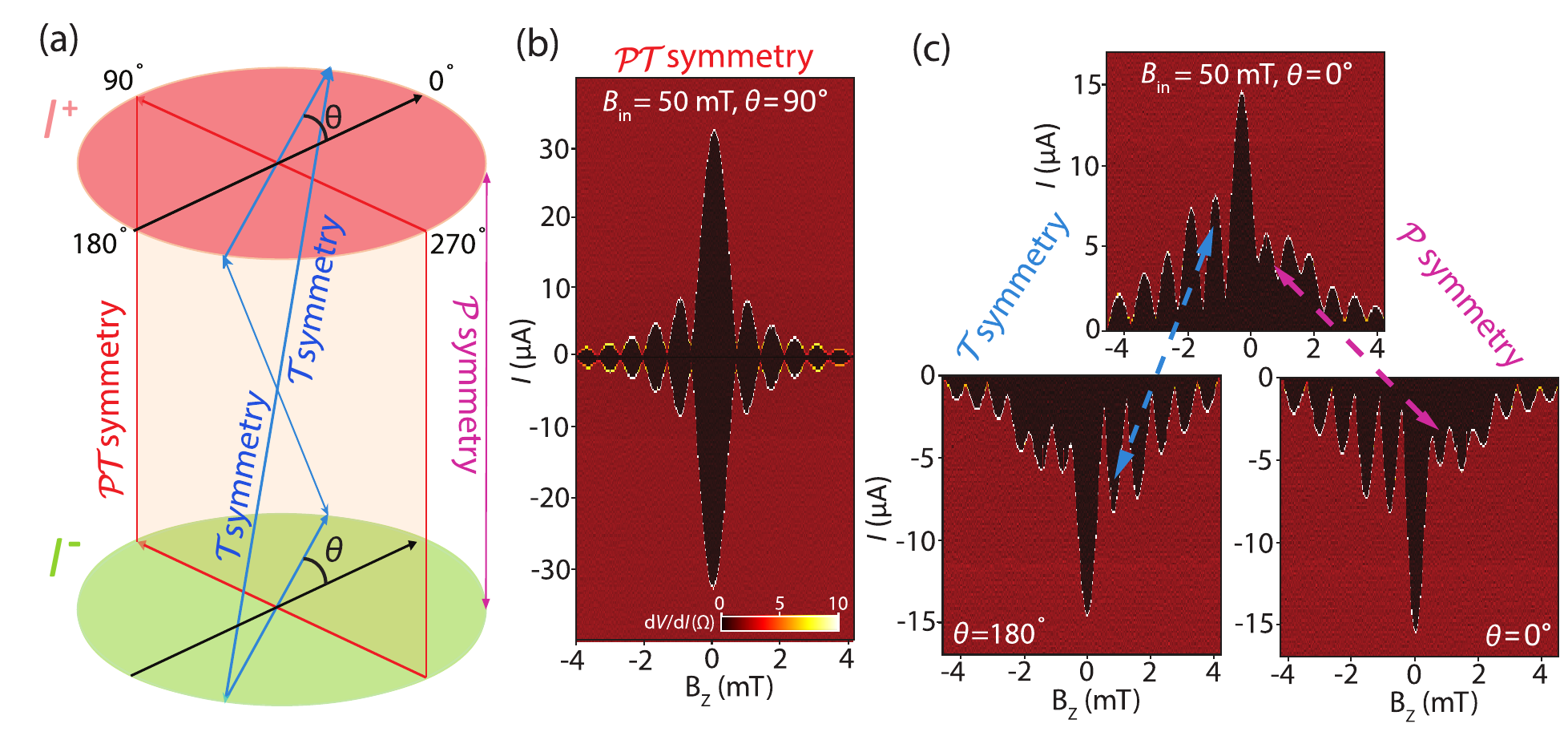} 
\caption{ \textbf{Field-current symmetry map of Josephson interference.}
(a) Field-current symmetry map constructed from the supercurrent interference patterns. The two in-plane-field components define continuous field coordinates, whereas $I^{+}$ and $I^{-}$ form two discrete current sectors. The configurations at $\theta=90^\circ$ and $270^\circ$ define $\mathcal{PT}$-invariant boundaries. Away from these boundaries, $\mathcal{PT}$ symmetry is broken, while symmetry-related configurations remain connected by the generalized $\mathcal{T}$ operation.
(b) Interference patterns measured along a $\mathcal{PT}$-invariant boundary. A symmetric Fraunhofer response is recovered over a broad range of in-plane fields despite a finite $B_{\rm in}$.
(c) Representative off-axis configurations measured at $B_{\rm in}=\SI{50}{\milli\tesla}$ and $\theta=0^\circ$ and $180^\circ$. The corresponding interference patterns illustrate the relation between symmetry-broken sectors connected by the $\mathcal{P}$ and generalized $\mathcal{T}$ operations. All figures share the same color scale.
}
\label{fig:Fig4}
\end{figure}

\section*{Discussion}

The agreement between experiment and the microscopic model indicates that the pronounced asymmetry is unlikely to originate from the intrinsic crystal symmetry of PtTe$_2$.
This distinguishes the observed response from asymmetric Josephson effects proposed for materials breaking inversion symmetry~\cite{KTLawPRB:2018AJE}. Bulk 1T-PtTe$_2$ is centrosymmetric, and a mechanism dominated by Rashba spin-orbit coupling and disorder~\cite{AssoulineNatComm:2019,Mason:2021JPCM} is unlikely to account for our observations. In particular, sufficiently strong spin-orbit or Zeeman coupling rotates the principal axis of maximal asymmetry away from the junction direction by approximately $90^\circ$~\cite{FlensbergPRB:2017_AnomalousFraunhoferPattern,AssoulineNatComm:2019}
(see also Fig.~\ref{fig:asym_vargfac} and SI Sec.~\ref{Sec:microscopic}), in contrast to the experimentally observed angular dependence.

Although spatial variations in the proximity effect or other local superconducting properties may contribute to the inhomogeneous phase landscape, the minimal model here shows that onsite disorder and flux dipoles are sufficient to reproduce the asymmetric interference without requiring fine-tuned disorder or dipole profiles. Similar symmetry and angular responses in NiTe$_2$ JJs further support the generic nature of this mechanism (SI Sec.~\ref{sec:NiTe2}). 

At the device level, these symmetry operations relate distinct experimentally controlled configurations, even though the microscopic Hamiltonian for a fixed disorder realization and magnetic-field configuration need not possess the corresponding exact symmetries. The field-current configuration therefore provides a symmetry-selective protocol for interrogating the Josephson interference pattern.
At fixed $B_{\rm in}$ and $\theta$, comparing the positive- and negative-current branches probes the device-level $\mathcal{P}$ symmetry, as illustrated for representative configurations in the insets of Fig.~\ref{fig:Fig1}(b). The generalized $\mathcal{T}$ symmetry is instead tested by simultaneously reversing the current and the complete magnetic-field configuration, namely $B_{z}\rightarrow-B_{z}$ and $\theta\rightarrow\theta+180^\circ$. Most notably, selecting $\theta=90^\circ$ or $270^\circ$ suppresses the flux-dipole-induced phase modulation and restores the $\mathcal{PT}$-symmetric Fraunhofer response, within experimental resolution, even at finite $B_{\rm in}$. Rotation of the in-plane field therefore switches the junction reversibly between symmetry-broken and symmetry-restored interference configurations, rather than merely diagnosing an existing symmetry state.

We summarize these symmetry operations in the field-current symmetry map shown in Fig.~\ref{fig:Fig4}(a), where the positive- and negative-current branches form distinct sectors parameterized by the in-plane-field orientation. The configurations at $\theta=90^\circ$ and $270^\circ$ define $\mathcal{PT}$-invariant boundaries, along which a symmetric Fraunhofer pattern is recovered over a broad range of in-plane fields, as shown in Fig.~\ref{fig:Fig4}(b). Away from these boundaries, the interference pattern becomes asymmetric, while configurations related by simultaneous reversal of the current and magnetic-field directions remain connected by the generalized $\mathcal{T}$ operation. Figure~\ref{fig:Fig4}(c) shows representative off-axis configurations at $B_{\rm in}=\SI{50}{\milli\tesla}$ and $\theta=0^\circ$ and $180^\circ$. The map therefore provides a unified representation of how the field-current geometry selects the symmetry operation being probed and determines whether the interference response is symmetry-broken or symmetry-restored.
 
These results establish a reconfigurable, symmetry-selective Josephson interferometer in which the magnetic-field orientation serves as a control parameter and the side-lobe asymmetry provides a phase-sensitive readout. The same device can selectively probe $\mathcal{P}$ and generalized $\mathcal{T}$ symmetries, while specific field orientations restore the $\mathcal{PT}$-symmetric response despite a substantial applied in-plane field. Beyond identifying the microscopic origin of the asymmetric interference, this capability suggests opportunities for vector magnetic-field sensing, disorder-sensitive imaging of supercurrent paths, and field-configurable nonreciprocal superconducting transport. More generally, it shows that flux focusing and disorder can be used as controllable ingredients for engineering the symmetry of mesoscopic superconducting interference.

\section*{Methods} 

\subsection*{Fabrication of JJs and measurements}
To investigate magnetic-field-dependent Josephson behavior, we fabricated PtTe$_2$-based JJs in the 1T phase, exfoliated onto a \SI{280}{\nano\meter} SiO$_2$/Si substrate. The PtTe$_2$ flakes used in this work have thicknesses ranging from \SIrange{20}{40}{\nano\meter}. Superconducting electrodes were formed using \SI{70}{\nano\meter} of sputtered NbTi, chosen for its high critical temperature and strong resilience to in-plane and out-of-plane magnetic fields. Full device geometries and fabrication details are provided in SI Sec.~\ref{sec:Device}.

All experiments were carried out at a base temperature of \(\approx\) \SI{50}{\milli\kelvin}. A current bias was applied across the junction, and the resulting voltage was measured [Fig.~\ref{fig:Fig1}(a)]. Magnetic fields were applied using a three-axis vector magnet, enabling independent control of the out-of-plane and in-plane field components. We denote the out-of-plane magnetic field as \(B_z\), which is calibrated as 
detailed in SI Sec.~\ref{sec:Device}. 
For angle-dependent measurements, an in-plane magnetic field \(B_{\rm in}\) was applied within the sample plane and rotated by an angle \(\theta\), measured relative to the junction current direction. In this convention, \(\theta = 0^\circ\) corresponds to \(B_{\rm in}\) parallel to the current, while \(\theta = 90^\circ\) corresponds to \(B_{\rm in}\) transverse to the current.

\subsection*{Calculation of the supercurrent and interference patterns } 
To capture the microscopic origin of the observed lobe asymmetry, we model the system as a two-dimensional SNS junction, where the normal region is described by a square lattice and sandwiched between conventional $s$-wave superconductors with a phase difference $\varphi$; we do not include the third spatial dimension. 
For the numerical calculations presented in the main text, the normal region is discretized on an $N_x \times N_y = 30 \times 120$ lattice with discretization lattice constant $a_0 = 3 \,\mathrm{nm}$. 
This corresponds to a simulation junction area $A^{\rm sim} = L \times W = 90 \times 360\,\mathrm{nm}^2$ ($L = N_x a_0$ and $W = N_y a_0$) and an aspect ratio $L/W = 1/4$ close to the device geometry. We also check that the main features are robust for a larger lattice size;  see Fig.~\ref{fig:asym_largeN} in SI Sec.~\ref{Sec:microscopic}.

Because the simulations employ a smaller junction area $A^{\rm sim}$ than the real devices, we define an effective field strength $B_{\rm in}^{\rm eff} = B_{\rm in}^{\rm sim} A^{\rm sim} / A$, which would generate the same total flux in a device of realistic size $A$. 
Using this scaling with the device parameter, $A = 2.2 \, \mu {\rm m}^2$, we obtain $B_{\rm in}^{\rm eff} \approx \SIrange{7}{63}{\milli\tesla}$ in our analysis, comparable to the experimental range, $B_{\rm in} = \SIrange{5}{70}{\milli\tesla}$.

The normal region includes   nearest-neighbor hopping, orbital effects, Zeeman coupling, Rashba spin-orbit interaction, and an onsite disorder potential with random strength uniformly distributed in the interval $[-W_{\rm dis}/2, W_{\rm dis}/2]$.  
We include a spatial variation of the $z$ component of the magnetic field within the normal region, with a dipole-like profile arising from the in-plane-field flux-focusing effect described in the main text. 
The flux-focusing effect associated with the applied perpendicular field plays no role and is not included here~\cite{Paajaste:2015ACN, FlensbergPRB:2017_AnomalousFraunhoferPattern} (see SI Sec.~\ref{Sec:microscopic}). 
Within this mechanism, the asymmetry is governed primarily by the dipole-induced phase $\Delta\phi_{z}$, which is proportional to $B_{\text{in}}$ and the effective area over which the flux dipole forms. 

The supercurrent, $I_{s} (\varphi) = (2e/\hbar) \frac{\partial F}{\partial \varphi}$, is obtained from the phase derivative of the free energy $F$, which we obtain from perturbative expansion in terms of the NS tunneling amplitude. We retain the leading contribution to the supercurrent from the fourth-order perturbation.
The supercurrent is then evaluated using the recursive Green-function method~\cite{diez2023symmetry,KTLawPRB:2018AJE, FurusakiPhysicaB:1994DirtyJJ,AsanoPRB:2002} with open boundary conditions. 
All simulations are performed at a temperature of $T=50~\mathrm{mK}$ as in the experiments with a superconducting gap $\Delta=0.9\,\mathrm{meV}$.

\subsection*{Parameters adopted in the numerical simulations}

The material parameters adopted in the simulations are partially guided by previous quantum-oscillation and Hall measurements on PtTe$_2$~\cite{Dongzhi2018:PtTe-QuantumOscil,Zheng2018:PtTe2-FS,Pavlosiuk2018:PtTe-Galvanomagnetic,2018PRBSingh:PtTe2-dHvA}. We use a hopping amplitude $t=25~\mathrm{meV}$, a chemical potential $\mu=70~\mathrm{meV}$ ($\mu/t\simeq3.0$), and a disorder strength $W_{\rm dis}/t=1.95$.
We adopt this disorder strength in the main text because it is in good agreement with the experimental observations. We further verify that the main features of the asymmetric interference pattern remain robust for the weaker disorder strength, as shown in Figs.~\ref{fig:asym_varfB}--\ref{fig:asym_largeN} in SI Sec.~\ref{Sec:microscopic}. 

We include Rashba spin-orbit and Zeeman terms, with weak coupling strengths chosen to reproduce the experimentally observed angular dependence. For stronger spin-orbit coupling, the principal axis of the asymmetry rotates away from the $0^\circ$--$180^\circ$ junction axis, consistent with previous studies of systems with strong spin-orbit interaction~\cite{FlensbergPRB:2017_AnomalousFraunhoferPattern,AssoulineNatComm:2019}, but in contrast to our observations. 
Increasing the Zeeman coupling produces a similar rotation of the principal axis. At larger in-plane fields, the calculated asymmetry evolves nonlinearly and its principal axis rotates away from the $0^\circ$--$180^\circ$ junction axis, which we attribute primarily to the enhanced Zeeman contribution. Consistently, increasing the $g$-factor at fixed field strength produces the same pronounced rotation in the simulations (Fig.~\ref{fig:asym_vargfac} in SI Sec.~\ref{Sec:microscopic}).
These results suggest that the effective low-field Zeeman coupling relevant to the junction is smaller than that inferred from the large values $g\sim10$--$100$ reported in high-field measurements of bulk PtTe$_2$~\cite{ParkinCommPhys:2024}.

The experiments further show that either the $B_z>0$ or the $B_z<0$ side of the interference pattern can be more strongly suppressed, depending on the device. Within a given device, however, the relative lobe heights on the two sides remain unchanged as $B_{\rm in}$ is varied for $|\theta|<90^\circ$. This device-dependent sign of the asymmetry can be naturally attributed to different microscopic disorder configurations. In Fig.~\ref{fig:Fig3}, we therefore present the experimental and simulated patterns as obtained, without relabelling the field direction or artificially adjusting the relative lobe heights. Consequently, the relative lobe heights in Fig.~\ref{fig:Fig3}(a) are opposite to those in Fig.~\ref{fig:Fig3}(c).

To clarify the role of the disorder configuration, we also calculate the interference patterns after applying mirror reflections to the simulated disorder profile. Reflection about the junction direction (the $x$ axis) reverses the interference pattern with respect to $B_z$, whereas reflection about the $y$ axis leaves it essentially unchanged (Fig.~\ref{fig:Frhfr_flipV} in SI Sec.~\ref{Sec:microscopic}). 
The results demonstrate that the sign of the lobe imbalance is determined by the spatial configuration of the disorder, while its systematic dependence on the in-plane-field orientation is governed by the flux-dipole mechanism.

\section*{Data availability}
The data supporting the findings of this study are openly available in Zenodo at \url{https://doi.org/10.5281/zenodo.23252653} (Ref.~\cite{zenodo_data}).

\section*{Acknowledgment}

We thank G.~Finkelstein, K.~T.~Law,  A.~Levchenko, A. Matos-Abiague, and N.-C.~Yeh for insightful discussions. Y.C.T. and C.T.K. would like to acknowledge funding support from the iMATEs program, Academia Sinica (AS), Taiwan with Grant No.~AS-iMATE-112-12, and the National Science and Technology Council (NSTC), Taiwan with Grant Nos.~NSTC-114-2112-M-001-053, NSTC-115-2112-M-001-050.  C.H.H. acknowledges support from NSTC, Taiwan with Grant Nos.~NSTC-114-2112-M-001-057, NSTC-114-2811-M-001-051, and 115-2112-M-001-024, and AS, Taiwan with Grant No.~AS-iMATE-114-12. C.N.K. acknowledges support from NSTC of Taiwan under Grant Nos. NSTC-113-2112-M-006-009-MY2   and  NSTC-115-2112-M-006-032-MY3.

\section*{Competing interests}
The authors declare no conflict of interest.

\putbib[Ref.bib]
\let\addcontentsline\temp 
\end{bibunit}


\clearpage
\bigskip 

\begin{bibunit}[unsrt]

\renewcommand\Authfont{\normalsize}
\renewcommand\Affilfont{\footnotesize}
\setlength{\affilsep}{0.3em}

\setcounter{equation}{0}
\setcounter{figure}{0}
\setcounter{table}{0}
\setcounter{page}{1}

\renewcommand{\theequation}{S\arabic{equation}}
\renewcommand{\thefigure}{S\arabic{figure}}
\renewcommand{\thetable}{S\arabic{table}}

\renewcommand{\citenumfont}[1]{S#1}
\renewcommand{\bibnumfmt}[1]{[S#1]}

\newcommand{\nn}{\nonumber \\}
\newcommand{\im}{\mathrm{Im}}
\newcommand{\re}{\mathrm{Re}}

\begin{center}
\large{\Large \textbf{Supplementary Information for ``Field-controlled breaking and restoration of parity-time symmetry in Josephson interference''}}\\
\vspace{15pt}
\fontsize{10}{12}
Yi-Chen~Tsai$^{1,2,3}$,~Yung-Yeh~Chang$^{1}$,~Tao-Yi~Hsu$^{1,4}$,~Thomas~Kuo$^{1,4}$,~Chia-Nung~Kuo$^{5,6,7}$,~Chin-Shan~Lue$^{5,6,7}$,~Kuei-Lin~Chiu$^{4,8}$,~Chen-Hsuan~Hsu$^{1,9,*}$,~and~Chung-Ting~Ke$^{1,10,*}$\\

$^{1}$Institute of Physics, Academia Sinica, Taipei 115201, Taiwan\\
$^{2}$Department of Engineering and System Science, National Tsing Hua University, Hsinchu 300044, Taiwan\\
$^{3}$Nanoscience and Technology Program, Taiwan International Graduate Program, Academia Sinica, Taipei 115201, Taiwan\\
$^{4}$Department of Physics, National Sun Yat-Sen University, Kaohsiung 804, Taiwan\\
$^{5}$Program on Key Materials, Academy of Innovative Semiconductor and Sustainable Manufacturing (AISSM), National Cheng Kung University, Tainan 70101, Taiwan\\
$^{6}$Department of Physics, National Cheng Kung University, Tainan 701, Taiwan\\
$^{7}$Taiwan Consortium of Emergent Crystalline Materials (TCECM), National Science and Technology Council, Taipei 10601, Taiwan\\
$^{8}$Research Center for Quantum Computing and Quantum Materials, National Sun Yat-Sen University, Kaohsiung 80424, Taiwan\\
$^{9}$Physics Division, National Center for Theoretical Sciences, Taipei 106319, Taiwan\\
$^{10}$Research Center for Critical Issues, Tainan 711010, Taiwan\\
$^{*}$Corresponding authors: chenhsuan@as.edu.tw, ctke@as.edu.tw
\end{center}

\tableofcontents

\clearpage

\section{Details about the fabrication and calibration }
\label{sec:Device}

\subsection{Device fabrication}

The 1T-PtTe$_2$ flakes were mechanically exfoliated inside the glove box to prevent oxidation. To minimize ambient exposure, a layer of PMMA was spin-coated immediately after exfoliation inside the glove box. The samples were then taken out and inspected under an optical microscope to identify suitable flakes. Device fabrication began with a first electron-beam lithography (EBL) step to define metallic alignment markers, followed by metal deposition and lift-off. After the markers were completed, the samples were immediately spin-coated with PMMA for the second EBL step, which defined the Josephson junction geometry. Before metal deposition, an O$_2$ plasma descum (\SI{30}{\second}, \SI{40}{\watt}) was applied to remove resist residues from the exposed PtTe$_2$ surface. Superconducting contacts were deposited by DC sputtering of \SI{70}{\nano\meter} NbTi, followed by lift-off in acetone to complete the junction fabrication. Following device fabrication, atomic force microscopy (AFM) measurements were performed on regions of the PtTe$_2$ flake not covered by electrodes to determine the actual thickness of the flake, which was found to be \SIrange{20}{40}{\nano\meter}.

\subsection{Coordinate calibration of the perpendicular magnetic field}

\begin{figure}[htbp]
\centering
\includegraphics[width=0.7\linewidth]{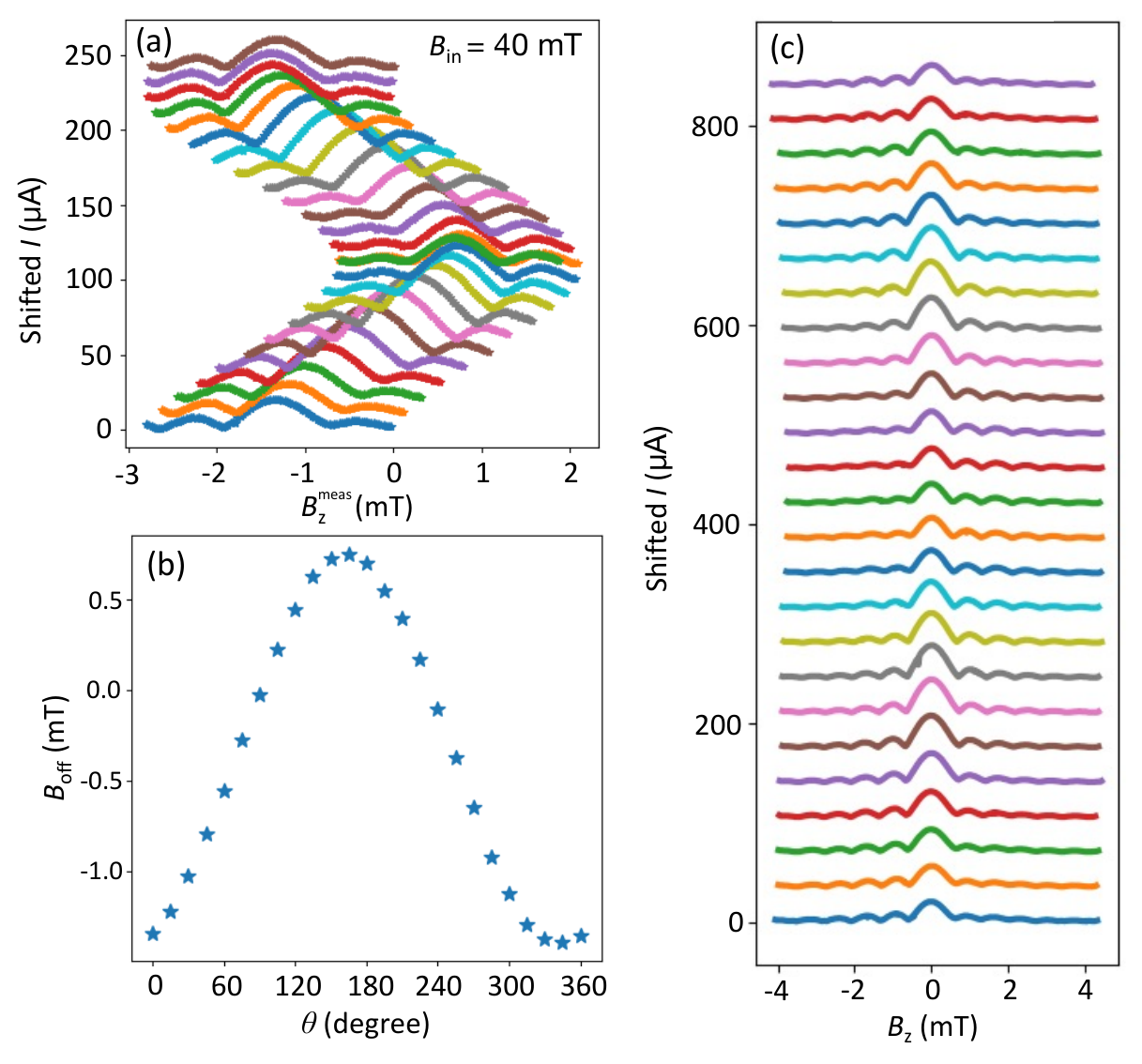}
\caption{
Angle-dependent calibration of the perpendicular magnetic field.
(a) Vertically shifted switching-current traces measured as a function of the nominal out-of-plane magnetic field \(B_z\) under a fixed in-plane field magnitude $B_\text{in}$=\SI{40}{\mt}.
(b) Angle-dependent field offset $B_\text{off}$, extracted from Gaussian fits to the dominant central maximum of each trace. (c) Corrected traces plotted as a function of
$B_z = B_z^{\mathrm{meas}} - B_{\mathrm{off}}$,
showing that the central maxima are aligned at $B_z=0$.
}
\label{Fig:S1}
\end{figure}

To align the sample coordinate with the vector magnetic field, we use the supercurrent interference patterns as a reference point to calibrate coordinates. An example of $B_\text{in}$=\SI{40}{\mt} is shown in Fig.~\ref{Fig:S1}(a). At each nominal rotation angle, the switching current \(I_\text{c}\) was measured as a function of the perpendicular magnetic field $B_z$. To account for residual misalignment of the vector magnet or offset due to sample mounting, which can shift the central \(I_\text{c}\) maximum away from the nominal \(B_z = 0\), we determined the field offset independently for each angle of the in-plane magnetic field. Specifically, the dominant central \(I_\text{c}\) maximum was identified in each trace, and then a Gaussian fit was made within a \(\pm 50\)-point window around this maximum. The fitted peak center was defined as the angle-dependent field offset, \(B_{\mathrm{off}}\); see Fig.~\ref{Fig:S1}(b). The field axis of each trace was then corrected according to
\[
B_z = B_z^{\mathrm{meas}} - B_{\mathrm{off}}.
\]
Fig.~\ref{Fig:S1}(c) shows that the central maxima of all traces were aligned at \(B_z\) = 0, enabling direct comparison of the interference patterns at different rotation angles. This procedure removes the residual perpendicular-field component induced by angular misalignment and establishes a common field reference for the full angle-dependent data set.

\FloatBarrier

\section{Additional experimental data for PtTe$_\text{2}$ junctions}
\label{sec:additional_data}

\subsection{Orthogonal junction orientations on the same PtTe$_\text{2}$ flake}

\begin{figure}[h]
\centering
\includegraphics[width=0.95\linewidth]{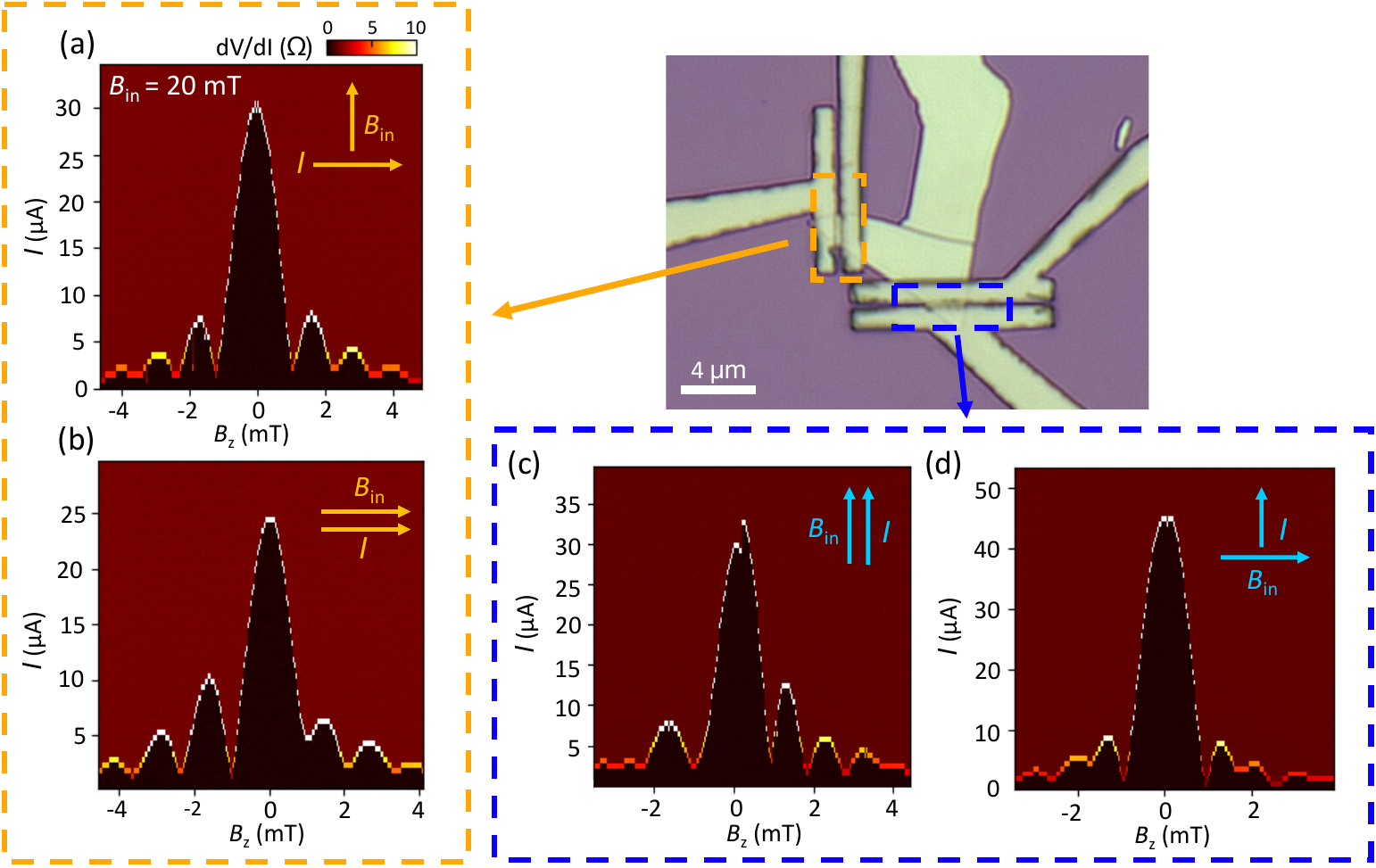}
\caption{Interference patterns of two orthogonally oriented Josephson junctions fabricated on the same PtTe$_2$ flake. The orange (a,b) and blue (c,d) dashed boxes in the optical image mark the two junctions, whose current directions are mutually perpendicular. 
For each junction, the interference pattern was measured at $B_\text{in}=\SI{20}{\milli\tesla}$ with the in-plane field oriented either parallel (b,c) or perpendicular (a,d) to the corresponding current direction. 
Because the two junctions are orthogonally oriented, a given global in-plane field direction is parallel to the current in one junction and perpendicular to that in the other.
For both junctions, the interference pattern becomes strongly asymmetric when $B_\text{in}$ is parallel to the corresponding current direction, whereas a nearly conventional, symmetric Fraunhofer pattern is recovered when $B_\text{in}$ is perpendicular to the current.
Because the two junctions are fabricated on the same PtTe$_2$ flake, this comparison shows that the Fraunhofer asymmetry is governed primarily by the relative angle between the in-plane field and the junction current, rather than by a fixed in-plane direction or a simple intrinsic anisotropy of the flake. All figures share the same color scale. 
}
\label{Fig:Orthogonal-JJ} 
\end{figure}

To rule out sample-to-sample variations and flake-intrinsic anisotropy, we compared two Josephson junctions fabricated on the same $\mathrm{PtTe}_2$ flake but with mutually perpendicular current directions, as indicated by the dashed boxes in the optical image of Fig.~\ref{Fig:Orthogonal-JJ}. Since the two junctions share the same flake, they also share the same crystallographic axes and intrinsic material properties. Therefore, if the observed interference asymmetry was mainly caused by an intrinsic in-plane anisotropy of the flake, such as an in-plane $g$-factor anisotropy~\cite{Mu_2021}, the response would be expected to depend on the orientation of the in-plane magnetic field relative to the principal crystallographic axes.

Nevertheless, as seen in Fig.~\ref{Fig:Orthogonal-JJ}, both orthogonal junctions show the same dependence on the relative field-current orientation: the interference pattern becomes strongly asymmetric when the in-plane field is parallel to the current, whereas a nearly conventional, symmetric Fraunhofer pattern is recovered when the field is perpendicular to the current. This demonstrates that the $\mathcal{PT}$-invariant orientation rotates with the junction rather than remaining fixed to a particular crystallographic axis of the flake. We therefore conclude that the asymmetry is governed primarily by the relative angle between the in-plane magnetic field and the junction current, rather than by intrinsic material anisotropy.

\FloatBarrier

\subsection{Supercurrent interference patterns for  $B_\text{in}=$~\SI{70}{\mt}}
\label{Sec:high-field}

\begin{figure}[!htbp]
\centering
\includegraphics[width=1\linewidth]{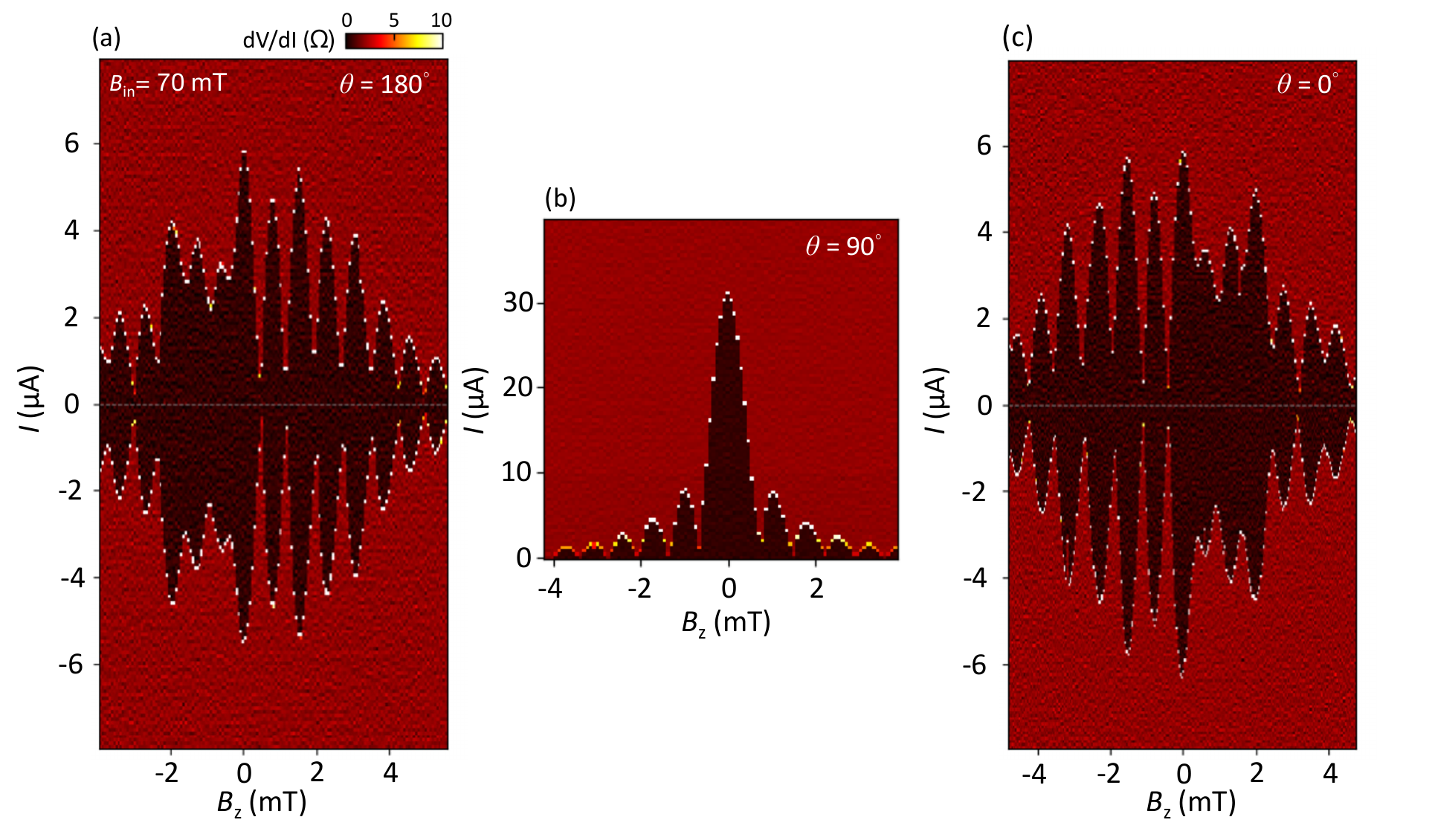}
\caption{Supercurrent interference patterns at a fixed in-plane magnetic field of $B_\text{in}=\SI{70}{\mt}$. The differential resistance $dV/dI$ is plotted as a function of bias current $I$ and perpendicular magnetic field $B_z$ for 
(a) $\theta=\ang{180}$, 
(b) $\theta=\ang{90}$, and 
(c) $\theta=\ang{0}$, where $\theta$ is measured relative to the current direction. For a parallel or antiparallel in-plane field, $\theta=\ang{0}$ or $\ang{180}$, the interference pattern becomes strongly asymmetric and develops a SQUID-like envelope. For the perpendicular orientation, $\theta=\ang{90}$, a nearly conventional, symmetric Fraunhofer pattern is recovered around $B_z=0$. The current-axis range in panel (b) differs from those in panels (a) and (c) to display the enhanced critical current.
All figures share the same color scale.}
\label{Fig:Interference-70mT}
\end{figure}

In this section, we present additional supercurrent interference data at $B_\text{in}=\SI{70}{\mt}$, including both the $I^{+}$ and $I^{-}$ branches, for $\theta=\ang{0}$, $\ang{90}$, and $\ang{180}$. Consistent with the results in the main text, the interference patterns become strongly asymmetric when the in-plane field is parallel or antiparallel to the current, corresponding to $\theta=\ang{0}$ and $\ang{180}$, as shown in Figs.~\ref{Fig:Interference-70mT}(a) and \ref{Fig:Interference-70mT}(c).
At this in-plane field, the interference pattern develops a SQUID-like envelope, reflecting a pronounced suppression of the central-lobe supercurrent. Nevertheless, the positive- and negative-current branches remain mutually symmetric.

By contrast, when the in-plane field is perpendicular to the current, $\theta=\ang{90}$, the interference pattern recovers a nearly conventional, symmetric Fraunhofer profile around $B_z=0$ [Fig.~\ref{Fig:Interference-70mT}(b)]. These results demonstrate the strong orientation dependence of the field-induced asymmetry, with the transverse in-plane-field direction acting as a symmetry-preserving axis of the junction. Increasing $B_\text{in}$ enhances the dipole-induced phase accumulation near the NS interfaces and thereby strengthens the asymmetry, whereas rotating the field to the transverse direction suppresses the dipole contribution and restores a symmetric response;   see Sec.~\ref{Sec:microscopic} for the details about the microscopic mechanism.

\FloatBarrier

\subsection{Continuous evolution of the interference patterns  }

In this section, we present additional supercurrent interference data obtained by measuring the critical current $I_\text{c}$ as a function of the perpendicular magnetic field $B_z$ while continuously varying $B_\text{in}$ over an extended field range. We compare the two orientations $\theta = 0^\circ$ and $\theta = 90^\circ$. 
 
\begin{figure}[H]
\centering
\includegraphics[width=0.95\textwidth]{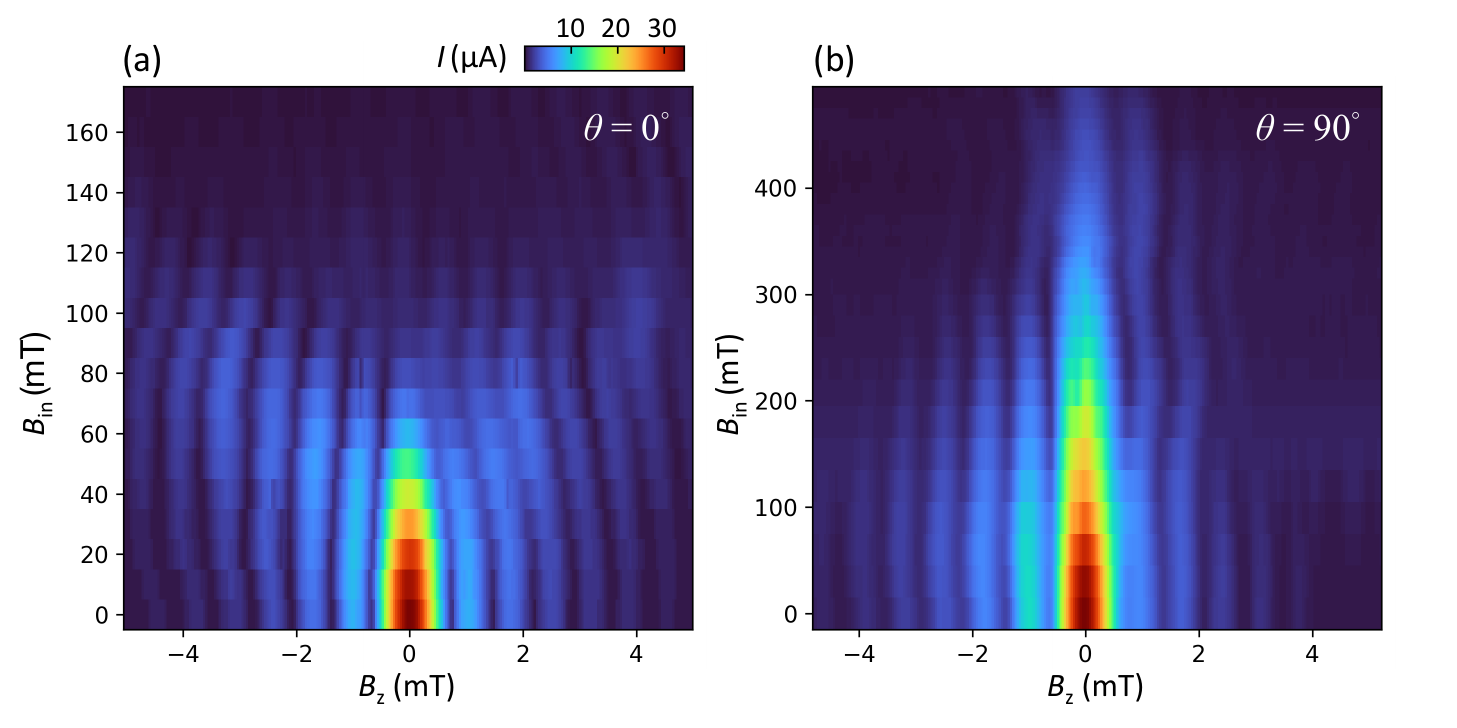}

\caption{Continuous evolution of the critical-current interference pattern with $B_\text{in}$. The critical current $I_\text{c}$ is plotted as a function of $B_\text{in}$ and the perpendicular field $B_z$. (a) For $\theta=\ang{0}$, increasing $B_\text{in}$ suppresses the central lobe and drives the pattern toward a SQUID-like profile. (b) For $\theta=\ang{90}$, the pattern remains nearly symmetric and Fraunhofer-like over a wider field range.
}
  \label{Fig:evolution}
\end{figure}

Figure~\ref{Fig:evolution}(a) shows the evolution of the interference pattern for $\theta = 0^\circ$. As $B_\text{in}$ increases, the Fraunhofer lobes become progressively distorted, the side-lobe asymmetry is enhanced, and the pattern increasingly deviates from the conventional symmetric Fraunhofer profile. At larger fields, it gradually evolves toward a SQUID-like pattern.

By contrast, when the in-plane field is perpendicular to the current, as shown in Fig.~\ref{Fig:evolution}(b), the central lobe remains robust over a substantially wider range of $B_\text{in}$, and the interference pattern retains a nearly symmetric Fraunhofer-like profile. For the $\theta=\ang{90}$ data, the slightly enhanced critical current on the negative-$B_z$ side may originate from sweep-direction-dependent vortex pinning, because $B_z$ was swept downward from \SI{500}{\milli\tesla} to \SI{0}{\milli\tesla}. This contribution can be removed by thermally cycling the sample above the critical temperature $T_\text{c}$ of the NbTi electrodes. The contrasting evolutions in Figs.~\ref{Fig:evolution}(a) and \ref{Fig:evolution}(b) indicate that an in-plane field parallel to the junction current perturbs the Josephson interference pattern much more strongly, consistent with the enhanced flux-dipole effect in this orientation, as discussed further in Sec.~\ref{Sec:microscopic}.

Previous studies have shown that proximity-induced Andreev pairs in the normal region can acquire a finite center-of-mass momentum~\cite{Hart2017,Chen2018,Flensberg:2016PRB}. Such finite-momentum pairing produces an additional directional phase gradient across the junction and has been reported in topological materials,
including HgTe and Bi$_\text{2}$Se$_\text{3}$, and in large spin-orbit/$g$-factor materials such as InSb and InAs~\cite{Hart2017,Chen2018,Ke2019,PhysRevB.108.094517}.
In the present device, however, the pronounced Fraunhofer-to-SQUID-like evolution in Fig.~\ref{Fig:evolution}(a), together with its strong dependence on the field-current orientation, is naturally explained by spatial phase modulation induced by the flux dipole even without strong spin-orbit and Zeeman couplings; see Sec.~\ref{Sec:microscopic} for the details about the microscopic mechanism.
Although a finite-momentum pairing contribution cannot be excluded, the observed evolution primarily supports the field-geometry-dependent flux-focusing mechanism.

\subsection{Symmetry probe across junctions of different channel lengths }

In addition to the main device with $L=\SI{0.8}{\micro\meter}$, we perform the same symmetry analysis on two longer junctions with $L=\SI{1.2}{\micro\meter}$ and $L=\SI{1.6}{\micro\meter}$.
The device dimensions are summarized in Table~\ref{tab:JJsummary}. The device dimensions were determined from optical microscope and AFM. 

\begin{figure}[!htbbp]
\centering
\includegraphics[width=0.95\linewidth]{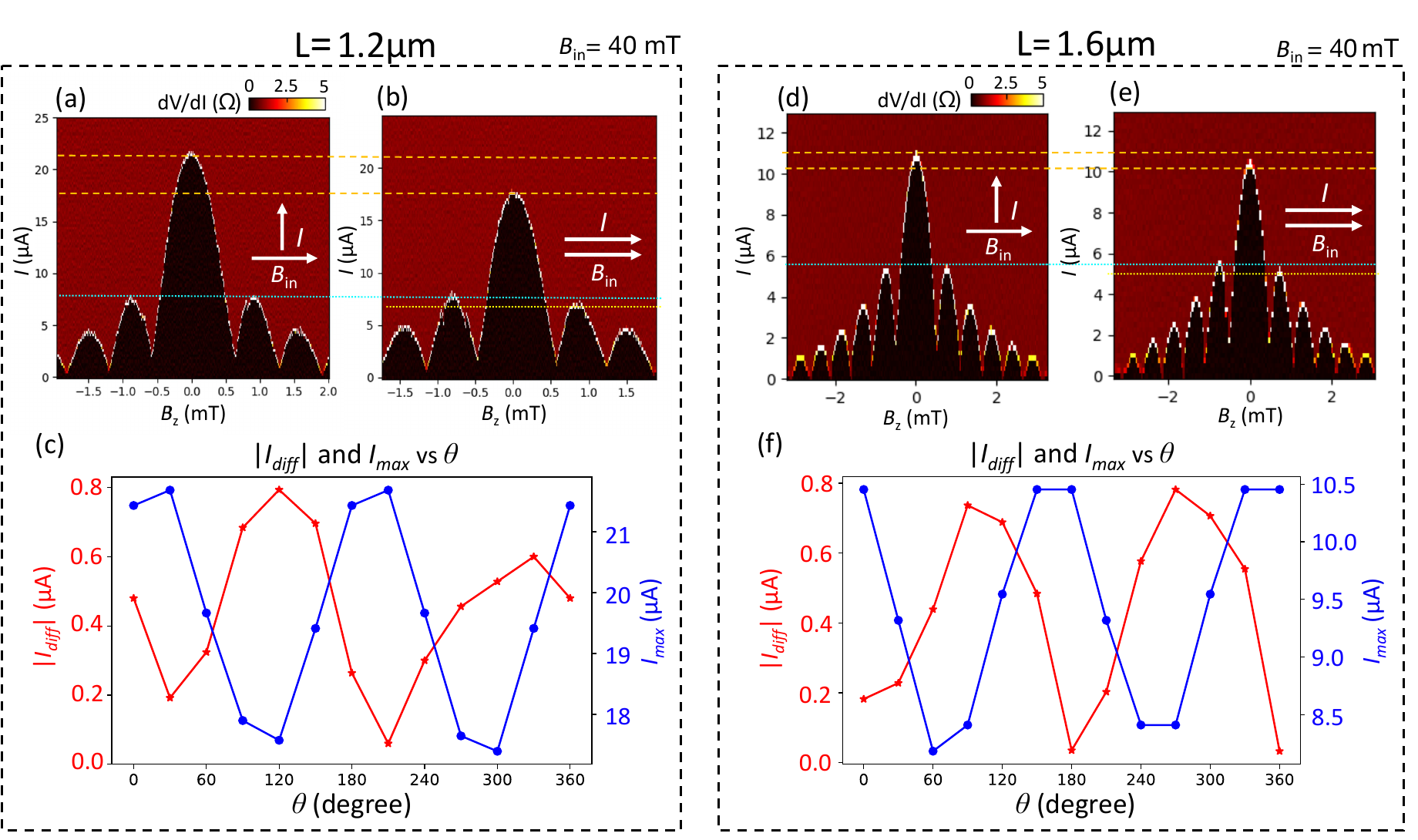}
\caption{Magnetic-field-dependent symmetry response of two longer PtTe$_2$ Josephson junctions at $B_\text{in} =\SI{40}{\mt}$.  
Differential-resistance maps for a junction with $L=\SI{1.2}{\micro\meter}$ with the in-plane field applied (a) perpendicular to the current direction and (b) along the current direction. (c) Angular dependence of the side-lobe asymmetry $\left|I_{\rm diff}\right|$ and maximum switching current for the same junction.
(d--f) Results analogous to (a)--(c), but for a junction with $L=\SI{1.6}{\micro\meter}$. 
 In both devices, the interference pattern is more asymmetric when the in-plane field is parallel to the current and becomes more symmetric when the field is perpendicular, consistent with behavior observed in the main device.}
\label{Fig:S2}
\end{figure}

As shown in Fig.~\ref{Fig:S2}, both devices exhibit the same characteristic angular dependence: the side-lobe asymmetry becomes more pronounced when the in-plane magnetic field is   parallel to the current direction, whereas the interference pattern becomes more symmetric when the in-plane field is perpendicular to the current. The extracted \(\left|I_{\rm diff}\right|\) and maximum switching current show the same two-fold modulation in angle, consistent with the behavior observed in the main device. These results indicate the universality of the observed symmetry response, which is not limited to a single junction length or a specific device geometry.

\begin{table}[!htbp]
\centering
\caption{Summary of the Josephson junction dimensions with different channel length $L$ and width $W$, along with their observed central-lobe current maximum  $I_{\mathrm{max}}$  and the maximum asymmetry $I_{\mathrm{diff}}$.}
\vspace{6pt}  
\begin{tabular}{c c c c c}
\hline
Device & Length $L$ (\si{\micro\meter}) & Width $W$ (\si{\micro\meter}) &  $I_{\mathrm{max}}$ (\si{\micro\ampere}) & Max $I_{\mathrm{diff}}$ (\si{\micro\ampere}) \\
\hline

JJ1 & 0.8 & 2.75 & 33.277 & 1.889\\
JJ2 & 1.2 & 2.5 & 21.681 & 0.794\\
JJ3 & 1.6 & 2.5 & 10.455 & 0.786\\
\hline
\end{tabular}
\label{tab:JJsummary}
\end{table}

\FloatBarrier

\section{Asymmetric interference pattern of  NiTe$_\text{2}$ junctions
}
\label{sec:NiTe2}

To determine whether the observed orientation dependence of the asymmetric interference patterns extends beyond PtTe$_2$, we examine a NiTe$_2$-based Josephson junction, where NiTe$_2$ is a related telluride semimetal often compared with PtTe$_{2}$ and PdTe$_{2}$~\cite{Xu2018NiTe2,PalNatPhys:2022}. We characterize the symmetry of its supercurrent interference pattern under different in-plane-field orientations.

\begin{figure}[H]
\centering
\includegraphics[width=0.5\linewidth]{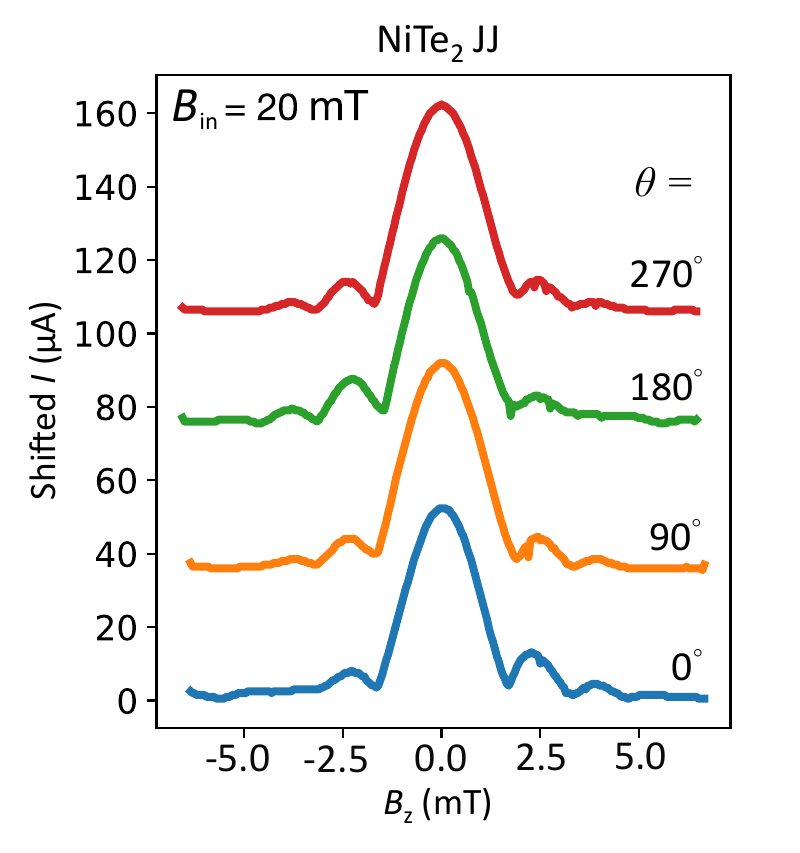}

\caption{Interference patterns of the NiTe$_2$ Josephson junction measured at a fixed in-plane magnetic field of $B_\text{in}$ =\SI{20}{\mt} for several field orientations. The pattern exhibits a pronounced orientation dependence: an asymmetric interference pattern is observed when the in-plane field is parallel or antiparallel to the current direction, whereas a symmetric Fraunhofer pattern is recovered when the field is perpendicular to the current.}
\label{Fig:NiTe2-JJ}
\end{figure}

As shown in Fig.~\ref{Fig:NiTe2-JJ}, the NiTe$_{2}$ junction exhibits the same orientation-dependent response as the PtTe$_{2}$ devices at $B_\text{in}=\SI{20}{\milli\tesla}$. The interference pattern remains nearly symmetric and Fraunhofer-like for an in-plane field with an angle with respect to current, $\theta=\ang{90}$ and $\ang{270}$, but becomes asymmetric for $\theta=\ang{0}$ and $\ang{180}$. This result demonstrates that the observed behavior is not specific to PtTe$_{2}$ and supports its broader relevance to related semimetallic Josephson junctions.

\FloatBarrier

\section{Microscopic mechanism of the interference   asymmetry }
\label{Sec:microscopic}

As introduced in the main text, at the device level the symmetry of the supercurrent interference pattern in the absence of in-plane fields is characterized through the relations~\cite{KTLawPRB:2018AJE,Kashiwaya:2019PRB}
\begin{eqnarray}
\mathcal{P}: && I^{\pm}(B_{z})=-I^{\mp}(B_{z}),
\label{Eq:P-symmetry}
\\
\mathcal{T}: && I^{\pm}(B_{z})=-I^{\mp}(-B_{z}),
\label{Eq:T-symmetry}
\end{eqnarray}
where $I^{\pm}$ denote the positive and negative current branches, respectively. Combining these relations, $\mathcal{PT}$ symmetry implies
\begin{equation}
I^{\pm}(B_{z})=I^{\pm}(-B_{z}),
\end{equation}
corresponding to a symmetric Fraunhofer pattern under reversal of the out-of-plane field.
Here, $\mathcal{P}$ and $\mathcal{T}$ denote operational symmetries of the measured device response, defined by transformations of the current branches and magnetic-field configuration. They do not require the microscopic Hamiltonian for a fixed disorder realization and magnetic-field configuration to possess exact inversion or time-reversal symmetry. In particular, a specific spatial disorder profile generally breaks microscopic inversion symmetry, whereas an applied magnetic field breaks microscopic time-reversal symmetry. The relations above instead connect experimentally accessible configurations and characterize whether the resulting Josephson interference patterns remain invariant under the corresponding operations.

In the following, we introduce a microscopic model incorporating disorder and flux dipoles generated by the in-plane field. Although the Hamiltonian for an individual disorder and field configuration need not obey exact $\mathcal{P}$ or $\mathcal{T}$ symmetry microscopically, the model reproduces the observed asymmetric interference and how the symmetry can be restored under changes in the field-current configuration.

\subsection{Model of the SNS junction and parameters}
\label{sec:model_parameter}

In this section, we introduce the model of the superconductor–normal metal–superconductor (SNS) Josephson junction used to calculate the supercurrent and its interference pattern. Although the experimental device is three-dimensional, its thickness is much smaller than its in-plane dimensions. The transverse modes along the $z$ direction are therefore quantized, allowing the electronic transport to be described effectively in two dimensions. This dimensional reduction also makes the numerical calculation more tractable.

\begin{figure}
    \centering
    \includegraphics[width=0.75\textwidth]{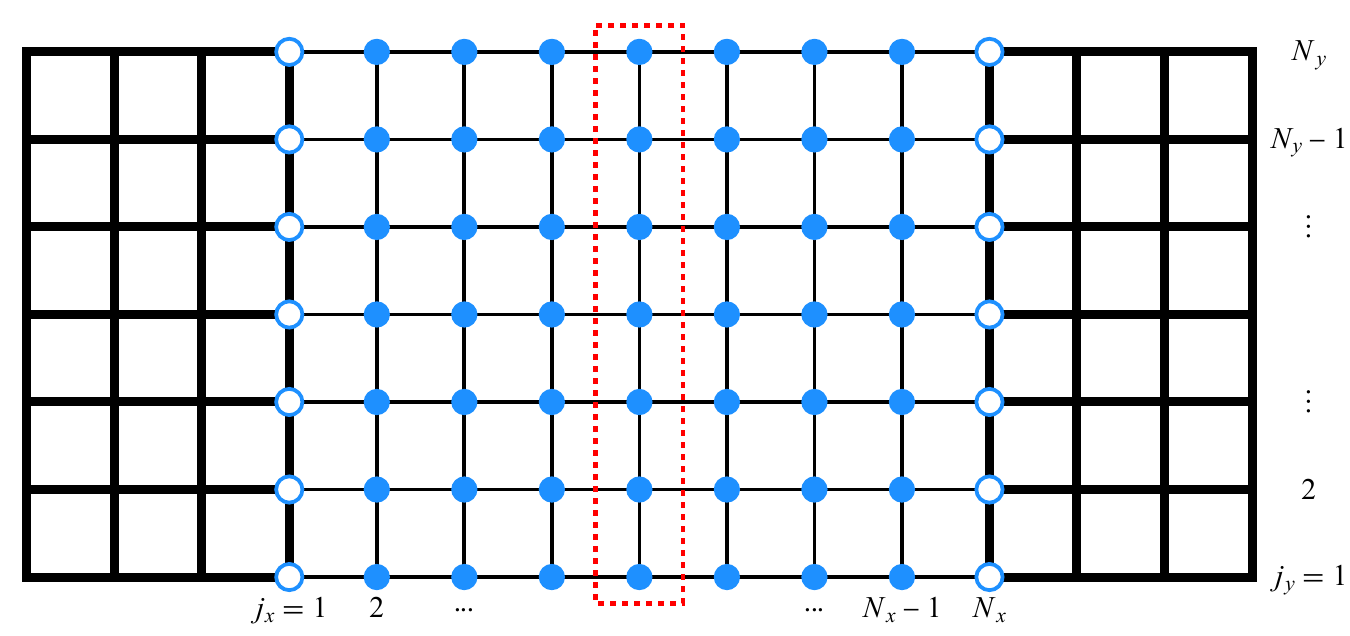}
    \caption{\textbf{Illustration of the discretized SNS junction.} The left and right regions with thick lines represent the superconducting leads. Blue solid circles in the middle region indicate lattice sites in the N region, whereas blue hollow circles indicate lattice sites at the NS interfaces. The red dashed rectangle highlights one of  the unit cells employed in the lattice recursive Green's-function calculations. This illustration is schematic and does not represent the actual numerical geometry.
    }
    \label{fig:schematic-SNS}
\end{figure}

We consider two conventional $s$-wave superconducting leads with a phase difference $\varphi$, separated by a disordered two-dimensional normal region of length $L$ and width $W$. The explicit Hamiltonian is given below. For the numerical calculations, the normal region is discretized on a square lattice of size $N_x \times N_y$ with lattice spacing $a_0$, where $N_x$ and $N_y$ denote the numbers of sites along the $x$ and $y$ directions, respectively. A schematic of the discretized SNS junction is shown in Fig.~\ref{fig:schematic-SNS}.

Throughout the simulations, we set $a_0 = 3$~nm for the numerical discretization lattice constant. 
To keep the computation tractable, we use junctions smaller than the experimental device while approximately preserving its aspect ratio. Most calculations are performed on a $30\times120$ lattice, corresponding to $L\times W=90\times360 \,{\rm nm}^2$. We also verify the results for a larger $45\times180$ lattice, corresponding to $L\times W=135\times540 \,{\rm nm}^2$, and find that the conclusions remain unchanged.

The supercurrent is taken to flow along the $+x$ direction, and the magnetic field is assumed to be nonzero only within the N region. We account for in-plane-field flux focusing at the NS interfaces, which distorts the in-plane magnetic flux and generates a spatially nonuniform, dipole-like out-of-plane field near the NS edges~\cite{FlensbergPRB:2017_AnomalousFraunhoferPattern}. For a simulated in-plane field of magnitude $B_{\mathrm{in}}^{\rm sim}$ oriented at an angle $\theta$ relative to the supercurrent, the induced out-of-plane field contributes $+f_B B_{\rm in}^{\rm sim} \cos\theta$ near one NS interface and $-f_B B_{\rm in}^{\rm sim} \cos\theta$ near the other. These contributions extend over a finite distance from the interfaces, as illustrated in Fig.~\ref{fig:Fig1} of the main text.
Consequently, the magnetic field becomes nonuniform along the $x$ direction, given  by
\begin{align}
    \bm{B} (\bm{r}) & = \left(  B_x (\bm{r}), ~  B_y (\bm{r}),~ B_z (\bm{r}) \right) \nn 
	& =\begin{cases}
\left(\sqrt{1-f_B^{2}}B_{\rm in}^{\rm sim} \cos(\theta),\quad B_{\rm in}^{\rm sim} \sin(\theta),\quad B_z+f_B B_{\rm in}^{\rm sim} \cos(\theta)\right), & 1\leq j_{x}\leq n_{x},\\
\left(B_{\rm in}^{\rm sim} \cos(\theta),\quad B_{\rm in}^{\rm sim} \sin(\theta),\quad B_z\right), & n_{x}\leq j_{x}\leq N_{x}-n_{x}+1,\\
\left(\sqrt{1-f_B^{2}}B_{\rm in} ^{\rm sim} \cos(\theta),\quad B_{\rm in}^{\rm sim} \sin(\theta),\quad B_z-f_B B_{\rm in}^{\rm sim} \cos(\theta)\right), & N_{x}-n_{x}+1\leq j_{x}\leq N_{x},
\end{cases}
\label{eq:full_B_dipole}
\end{align}
where $\bm{r} = (x,\, y)$, $n_x$ denotes the spatial extent over which flux focusing occurs, and the factor $0\leq f_B \leq 1$  characterizes the efficiency for which the in-plane field is converted into the out-of-plane field: $f_B = 0$  corresponds to no flux focusing, while $f_B = 1$ corresponds to the in-plane field being fully converted into a  out-of-plane component.
One can see that the $y$-component of the magnetic field remains unchanged in Eq.~\eqref{eq:full_B_dipole}, since the in-plane flux-focusing arises solely from the $x$-component of the magnetic field [see Figs.~\ref{fig:Fig1}(c) and \ref{fig:Fig1}(d) of the main text]. Flux focusing associated with $B_z$ is also present. Since it does not affect the lobe asymmetry or the Fraunhofer-to-SQUID evolution considered here, we do not include this contribution in the simulations. Nevertheless, out-of-plane flux focusing can produce aperiodic minima in the interference pattern~\cite{FlensbergPRB:2017_AnomalousFraunhoferPattern,Paajaste:2015ACN}.  

In the calculation, the gauge field from the external magnetic field is incorporated through the Peierls substitution, where the hopping from site $\bm{j}^\prime$ to $\bm{j}$ acquires a gauge-dependent phase,
\begin{align}
    \exp\left[\frac{ie}{\hbar}\int^{\bm{j}a_0}_{\bm{j}^{\prime}a_0} d\bm{r}\cdot\bm{A}(\bm{r})\right] \, ,\quad   \bm{B} = \nabla \times \bm{A} \,.
\end{align}
We adopt the Landau gauge for the vector potential $\bm{A}$, 
\begin{align}
    \bm{A} ({\bm r}) = \left( -yB_z (\bm{r})  , \,  -z B_{x}(\bm{r}), \,- x B_{y}(\bm{r})    \right)\, , 
    \label{eq:A-field_dipole}
\end{align}
with $B_{x,y,z} (\bm{r})$ given in Eq.~\eqref{eq:full_B_dipole}. The Peierl substitution along the $x$ direction can be written as 
\begin{align}
   \exp\left[-\frac{ie}{\hbar}\int_{j_{x}a}^{\left(j_{x}\pm1\right)a}y B_{z} (\bm{r})dx\right]  =\exp\left[\mp\frac{\pi ij_{y}\phi_{z}^{{\rm eff}}(j_{x})}{(N_{x}-1)(N_{y}-1)}\right]\,,
\end{align}
where $\phi_{z}^{\rm eff}$ denotes the effective magnetic flux (in units of the flux quantum $\Phi_0 \equiv h/2e$), 
\begin{align}
    \phi_{z}^{{\rm eff}}(j_{x})=\begin{cases}
\phi_{z}+\Delta\phi_{z}, & 1\leq j_{x}\leq n_{x},\\
\phi_{z}, & n_{x}\leq j_{x}\leq N_{x}-n_{x}+1,\\
\phi_{z}-\Delta\phi_{z}, & N_{x}-n_{x}+1\leq j_{x}\leq N_{x}\,.
\end{cases}
\end{align}
In the above, we define the flux $\phi_z$ arising from $B_z$ and the additional term $\Delta\phi_z$ from flux focusing as follows,  
\begin{align}
     \phi_{z}  \equiv \frac{LW B_z}{\Phi_{0}} \, , \quad 
     \Delta \phi_{z}  \equiv \frac{LW f_B B_{\rm in}^{\rm sim} \cos(\theta)}{\Phi_0}\, .
\end{align}
The Peierls phase corresponding to $y$-direction hopping is given by
\begin{align}
    \exp\left(-\frac{ie}{\hbar}\int^{j_y a}_{j_y^\prime  a}z B_{x} (\bm{r}) dy\right)\, ,
\end{align}
which, in our 2D model, does not generate any nontrivial phase factor.

The N region is described by a tight-binding model of a 2D disordered semiconductor, 
\begin{align}
H_{\rm N} &= H_0 + H_{\rm SOI} + H_Z + H_{\rm dis}\, ,
\end{align}
where $H_0$ is the kinetic term with the nearest-neighbor  hopping, $H_{\rm SOI}$ the spin–orbit interaction (SOI), $H_Z$ the isotropic Zeeman coupling, and $H_{\rm dis}$ the disorder potential.
Including the Peierls factor,   $H_0$ takes the form,
\begin{align}
H_{0}=\sum_{\bm{j},\bm{j}^{\prime},\sigma} & \Bigg\{\left(4t-\mu\right)\delta_{j_{x},j_{x}^{\prime}}\delta_{j_{y},j_{y}^{\prime}} \nn
& - t \exp\left[-\frac{i\pi j_{y}\phi^{\rm eff}_{z} (j_x)}{(N_{x}-1)(N_{y}-1)}\right] \delta_{j_{x}-1,j_{x}^{\prime}}\delta_{j_{y},j_{y}^{\prime}}- \exp\left[\frac{i\pi j_{y}\phi^{\rm eff}_{z} (j_x)}{(N_{x}-1)(N_{y}-1)}\right] \delta_{j_{x}+1,j_{x}^\prime}\delta_{j_{y},j_{y}^{\prime}}\nn
& -t  \left(\delta_{j_{x},j_{x}^{\prime}}\delta_{j_{y}-1,j_{y}^{\prime}}+\delta_{j_{x},j_{x}^{\prime}}\delta_{j_{y}+1,j_{y}^{\prime}}\right) \Bigg\} c_{\bm{j}\sigma}^{\dagger}c_{\bm{j}^{\prime}\sigma}\, ,
\end{align}
where $\bm{j} = (j_x,\,  j_y)$ with $j_x \in\{1,\cdots,N_x\}$ and $j_y \in\{1,\cdots,N_y\}$, $\sigma\in \{\uparrow,\downarrow\}$ denotes the spin indices, $t$ represents the hopping strength and $\mu$ is the chemical potential. 

In the numerical calculations, we model the N region using a single electron-like band with an isotropic Fermi surface. Although multiband effects may modify the detailed interference pattern, this minimal description captures the essential mechanism underlying the asymmetry. The electron-like dispersion fixes the sign convention relating the Peierls phases to the magnetic-field dependence of the supercurrent. Replacing it with a hole-like dispersion reverses this relation, corresponding to the transformation $B_z\rightarrow -B_z$ of the Fraunhofer pattern, while leaving the qualitative asymmetry unchanged. We note, however, that doped PtTe$_2$ hosts both electron- and hole-like Fermi pockets because its chemical potential lies substantially away from charge neutrality~\cite{MingzheNatCommPtTe2:2017}.

The SOI term $H_{\rm SOI}$ includes both Rashba and Dresselhaus contributions and is given by
\begin{align}
    H_{\rm SOI} = \sum_{\bm{j},\bm{j}^{\prime}} & \Bigg[\left(-\frac{\alpha_{R}}{2ia_{0}}\sigma_{x}+\frac{\alpha_{D}}{2ia_{0}}\sigma_{y}\right) \delta_{j_{x},j_{x}^{\prime}}\delta_{j_{y}+1,j_{y}^{\prime}}+\left(\frac{\alpha_{R}}{2ia_{0}}\sigma_{x}-\frac{\alpha_{D}}{2ia_{0}}\sigma_{y}\right)\delta_{j_{x},j_{x}^{\prime}}\delta_{j_{y}-1,j_{y}^{\prime}} \nn
    & + \exp\left[\frac{i\pi j_{y}\phi^{\rm eff}_{z} (j_x)}{(N_{x}-1)(N_{y}-1)}\right] \left(\frac{\alpha_{R}}{2ia_{0}}\sigma_{y}+\frac{\alpha_{D}}{2ia_{0}}\sigma_{x}\right)\delta_{j_{x}+1,j_{x}^{\prime}}\delta_{j_{y},j_{y}^{\prime}}\nn
    & + \exp\left[-\frac{i\pi j_{y}\phi^{\rm eff}_{z} (j_x)}{(N_{x}-1)(N_{y}-1)}\right] \left(-\frac{\alpha_{R}}{2ia_{0}}\sigma_{y}-\frac{\alpha_{D}}{2ia_{0}}\sigma_{x}\right)\delta_{j_{x}-1,j_{x}^{\prime}}\delta_{j_{y},j_{y}^{\prime}} \Bigg] c_{\bm{j}}^{\dagger}c_{\bm{j}^{\prime}} \, ,
    \label{eq:H_SOI}
\end{align}
with $\alpha_R$ and $\alpha_D$ being the Rashba and Dresselhaus coefficients, respectively, $\sigma_{x,y}$ being the Pauli matrices acting on the spin subspace, and $c_{\bm j} \equiv \left( c_{\bm j\uparrow}, \, c_{\bm j\downarrow}  \right)^T$ the spinor. In our numerical simulations, we will set $\alpha_D = 0$ as discussed below, but we keep the Dresselhaus term in Eq.~\eqref{eq:H_SOI} for completeness.

The isotropic Zeeman term is given by
\begin{align}
        H_{Z} = -\frac{1}{2} \sum_{\bm{j}} g \bm{B} \cdot \bm{\sigma}c^\dagger_{\bm{j}}c_{\bm{j}}\,,
\end{align}
with $g$ being the Land\'e $g$ factor. 

Disorder is modeled by a random onsite potential,
\begin{align}
    H_{\rm dis} = \sum_{\bm{j},\sigma} V_{\bm{j}} c^\dagger_{\bm{j}\sigma}c_{\bm{j}\sigma}\, ,
    \label{eq:Hdis}
\end{align}
where $V_{\bm{j}}$ is independently drawn from a uniform distribution over $[-W_{\rm dis}/2,\,W_{\rm dis}/2]$, with $W_{\rm dis}$ denoting the disorder strength.

\begin{sidewaystable}[p]
\centering

\caption{\textbf{Material parameters for PtTe$_2$.}
The Fermi-surface cross-sectional area $A_F$, Fermi wave vector $k_F$,
effective mass $m^*$ (in units of the free-electron mass $m_e$), and
quantum mobility $\mu_q$ are taken from independent quantum-oscillation
and Hall measurements~\cite{
Dongzhi2018:PtTe-QuantumOscil,
Zheng2018:PtTe2-FS,
Pavlosiuk2018:PtTe-Galvanomagnetic,
2018PRBSingh:PtTe2-dHvA}.
The estimated hopping amplitude $t$, chemical potential $\mu$, 
mean free path $l_e$, and disorder strength $W_{\rm dis}$ are listed.  
}

\label{tab:QuanOscil-para}

\footnotesize
\setlength{\tabcolsep}{4.5pt}
\renewcommand{\arraystretch}{1.2}

\begin{tabular}{cccccccccc}
\hline\hline

\shortstack{
$A_F$\\
($10^{-4}\,\text{\AA}^{-2}$)
}
&
\shortstack{
$k_F$\\
($10^{8}\,\mathrm{m}^{-1}$)
}
&
\shortstack{
$m^*/m_e$
}
&
\shortstack{
$\mu_q$\\
($10^3\,\mathrm{cm^2\,V^{-1}\,s^{-1}}$)
}
&
\shortstack{
$l_e$\\
(nm)
}
&
\shortstack{
$t$\\
(meV)
}
&
\shortstack{
$\mu$\\
(meV)
}
&
$\mu/t$
&
$W_{\rm dis}/t$
&
Ref. \\

\hline

98
& 5.6
& 0.2
& 9.6
& 356
& 21.2
& 59.7
& 2.8
& 0.45
& \cite{Dongzhi2018:PtTe-QuantumOscil}
\\

103
& 5.7
& 0.2
& 6.0
& 225
& 28.2
& 61.9
& 2.2
& 0.44
&
\\

232
& 8.6
& 0.3
& 3.8
& 218
& 14.6
& 93.9
& 6.4
& 1.3
&
\\

\hline

--
& --
& 0.11
& --
& --
& 38.5
& --
& --
& --
& \cite{Zheng2018:PtTe2-FS}
\\

--
& --
& 0.27
& --
& --
& 15.7
& --
& --
& --
&
\\

\hline

103
& 5.7
& 0.11
& 2.14
& 80.7
& 38.5
& 112.5
& 2.9
& 0.97
& \cite{Pavlosiuk2018:PtTe-Galvanomagnetic}
\\

234
& 8.6
& 0.21
& 2.04
& 115.9
& 20.2
& 134.2
& 6.6
& 1.84
&
\\

\hline

110
& 6.0
& 0.21
& 3.3
& 180
& 20.2
& 129$^{\mathrm{a}}$
& --
& --
& \cite{2018PRBSingh:PtTe2-dHvA}
\\

130
& 6.5
& 0.3
& --
& --
& 16.3
& 124$^{\mathrm{b}}$
& --
& --
&
\\

210
& 8.3
& 0.3
& --
& --
& 13.2
& 164$^{\mathrm{c}}$
& --
& --
&
\\

\hline\hline
\end{tabular}

\vspace{6pt}

\begin{minipage}{0.95\linewidth}
\footnotesize
$^{\mathrm{a,b,c}}$
The values of the chemical potential are taken from
Ref.~\cite{2018PRBSingh:PtTe2-dHvA}, where $\mu$ is evaluated from
$\mu=\hbar v_F k_F$ under the assumption of a linear-in-momentum
dispersion, rather than the parabolic dispersion adopted here.
\end{minipage}

\end{sidewaystable}

\textit{Parameter estimations.}--We estimate the relevant material parameters   for the subsequent  calculation, as  summarized in Table~\ref{tab:QuanOscil-para}. 
The hopping amplitude $t$ is estimated by matching the low-energy lattice dispersion to the continuum effective-mass description, namely
\begin{align}
t=\frac{\hbar^2}{2m^* a_0^2}\, .
\end{align}
The experimental values of $m^*/m_e \approx 0.1$--$0.3$ with the free-electron mass $m_e$ lead to
$
    t= 12-39\, \rm{meV},
$
which motivated us to take $t = 25~\mathrm{meV}$ throughout the simulations. 
We estimate the chemical potential $\mu$ from a parabolic-band approximation,
\begin{align}
\mu \simeq \frac{\hbar^2 k_F^2}{2m^*} ,
\end{align}
with Fermi wave vector $k_F$ obtained by assuming a circular cross section of the Fermi surface, $A_F=\pi k_F^2.$ The value of $A_F$ can   be extracted from the quantum oscillation, leading to  $ 
\mu/t \simeq 2.2 - 6.6  $, motivating our choice of $\mu/t =3$. 

For the disorder strength $W_{\rm dis}$, we first estimate the mean free path,
$l_e = \frac{\hbar \mu_q k_F}{e}  = 80.7 - 356\, \rm{nm}$, from the relation $l_e=v_F\tau$ and the mobility $\mu_q=e\tau/m^*$, with the Fermi velocity $v_F=\hbar k_F/m^*$  and transport scattering time $\tau$. 
Using the estimated mean free path $l_e$, the disorder strength can be inferred from~\cite{FlensbergPRB:2017_AnomalousFraunhoferPattern}
\begin{align}
   \frac{W_{\rm dis}}{t}
   =
   \left(\frac{48a_0}{l_e}\right)^{1/2}
   \left(\frac{\mu}{t}\right)^{1/4}
   =
   0.44\text{--}1.84.
\end{align}
This range motivates the choice $W_{\rm dis}/t=1.1$ as a representative value for intermediate disorder strength. We also consider the somewhat stronger value $W_{\rm dis}/t=1.95$, which yields particularly good agreement with the main experimental features.

Because bulk PtTe$_2$ is centrosymmetric, spin–orbit-induced band splitting is not expected, and the bands remain effectively spin degenerate. Our devices use relatively thick PtTe$_2$ flakes, in contrast to thinner samples where surface inversion asymmetry may become important. Moreover, first-principles calculations indicate that the Rashba contribution dominates over the Dresselhaus-type term~\cite{Sino2021:Pt-Janus}. We therefore include a weak Rashba spin–orbit coupling with $\alpha_R\sim0.1~\mathrm{eV\cdot}$\AA~and neglect the Dresselhaus-type contribution.

\subsection{Calculation of the supercurrent}

The supercurrent is computed by adopting the following formula, derived within fourth-order perturbation theory in the NS tunneling amplitude ~\cite{Flensberg:2016PRB,FlensbergPRB:2017_AnomalousFraunhoferPattern},
\begin{align}
    I_{s}^{(4)}(\varphi)
    &=\left(\frac{-4et}{\hbar}\right) 
    \sum_{n}\frac{\tilde{\kappa}^2\tilde{\Delta}^2 k_{B}\widetilde{T}}{\hbar\widetilde{\omega}_{n}^{2}+\tilde{\Delta}^{2}}
    \im \Bigg(
    e^{-i\varphi}\,
    \Tr\Big\{
    \bigl[D_z\,\widetilde{\mathbb{G}}^{RL}(i\omega_n)\,D_z\bigr]\,
    \bigl[P\,\widetilde{\mathbb{G}}^{RL}(-i\omega_n)\,P\bigr]^{\mathsf T}
    \Big\}
    \Bigg) \, ,
    \label{eq:Is-discretized_vect}
\end{align}
which is rescaled by $et/\hbar$, to give dimensionless quantity.  
In the above, the trace is taken over the lattice sites  along $y$ direction and spin indices, $\omega_n = (2n+1)\pi k_B T/\hbar$ is the  fermion Matsubara frequency, and $D_z$ and $P$ are matrices of dimension $2N_y \times 2N_y$, defined as follows:
\begin{align}
D_{z}=\begin{pmatrix}\sigma_{z}\\
 & \sigma_{z}\\
 &  & \ddots\\
 &  &  & \sigma_{z}
\end{pmatrix}_{N_{y}\times N_{y}},\quad P=\begin{pmatrix}\sigma_{x}\\
 & \sigma_{x}\\
 &  & \ddots\\
 &  &  & \sigma_{x}
\end{pmatrix}_{N_{y}\times N_{y}}.
\label{eq:def_Dz_P}
\end{align}
We include the other dimensionless energy scales,
\begin{align}
    \tilde{\Delta}&\equiv\frac{\Delta}{t},\quad \tilde{\kappa}\equiv \frac{\kappa}{t}=  \frac{\gamma^{2}\nu_{sc}\pi a_0^{3}}{t}, \quad 
    k_{B} \widetilde{T} \equiv\frac{k_{B}T}{t} , \quad 
   \hbar\widetilde{\omega}_{n} =(2n+1)\pi k_{B} \widetilde{T}  ,
\end{align}
where $\gamma$ denotes the NS tunneling amplitude and $\nu_{\rm sc}$ is the normal-state density of states of the superconductor. We also define the dimensionless Matsubara Green's function (matrix), 
\begin{align}
\widetilde{\mathbb{G}}^{RL} (i\omega_n) \equiv    \left\langle N_x \right|\frac{1}{i\hbar \widetilde{\omega}_{n} \mathbb{I}   -\tilde{H}_{\rm N}}\left| 1 \right\rangle\, , \quad \widetilde{H}_{\rm N} \equiv  H_{\rm N}/t  ,
\label{eq:G_dimensionless}
\end{align}
sandwiched by the rightmost ($j_x=N_x$) and leftmost ($j_x=1$) states of the N region.
Here, $\mathbb{I}$ represents the identity matrix. 
Equation~\eqref{eq:G_dimensionless} is evaluated using the recursive Green's function method~\cite{KTLawPRB:2018AJE, FurusakiPhysicaB:1994DirtyJJ,AsanoPRB:2002,diez2023symmetry} with open boundary conditions (see Fig.~\ref{fig:schematic-SNS}), where the recursive unit cell is highlighted by the red dashed rectangle. 

As a remark, the supercurrent in Eq.~\eqref{eq:Is-discretized_vect} contains only the first-harmonic contribution and can therefore always be written in the form, 
\begin{align}
    I^{(4)}_s(\varphi) = I \sin (\varphi - \varphi_0),
    \label{eq:Is4_sin}
\end{align}
with  the  critical current $I$ and phase offset $\varphi_0$.

\subsection{Additional numerical results of the side-lobe asymmetry}
\label{sec:asym_results}

In this section, we present additional numerical results for the interference patterns and their side-lobe asymmetry beyond those discussed in the main text.
Unless otherwise specified, we adopt the parameters in our calculations:   $t=25\,\mathrm{meV}$,   $\mu=70\,\mathrm{meV}$ ($\mu/t\simeq 3.0$), temperature $T=50\,\mathrm{mK}$, transparency $\kappa = 3$ meV, superconducting gap $\Delta=0.9\,\mathrm{meV}$, $\alpha_R \sim 0.1~\mathrm{eV\cdot}$\AA; see  Sec.~\ref{sec:model_parameter}
 for detailed estimation.

The side-lobe asymmetry is defined as~\cite{FlensbergPRB:2017_AnomalousFraunhoferPattern,Flensberg:2016PRB}
\begin{align}
A_n=\Bigg| \frac{I^{(-n)}-I^{(n)}}{I^{(n)}+I^{(-n)}}\Bigg| \, ,
\label{eq:asymm}
\end{align}
where $+n (-n)$ corresponds to the $n$-th side lobe to the right (left) of the central  lobe.  
We systematically examine the dependence of side-lobe asymmetries defined in Eq.~\eqref{eq:asymm} on various parameters, including flux-focusing factor $f_B$,  in-plane field strength $B_{\rm in}^{\rm sim}$, $n_x$, and $g$ factor. 
In particular, we examine how the magnitude  of the asymmetry and the principal axis of its angular dependence evolve. 

Unless otherwise stated, we use the disorder strength $W_{\rm dis}/t=1.95$ and the disorder realization denoted by Dis.~A for a lattice size of $30\times120$, as in Figs.~\ref{fig:Fig2}(e,f) and \ref{fig:Fig3}(c,d) of the main text, because they well reproduce the main experimental features. In Sec.~\ref{sec:asym_results_weaker}, we will examine the weaker disorder strength $W_{\rm dis}/t=1.1$, other disorder realizations as well as a different lattice size, showing that most qualitative features remain robust.

\begin{figure}[th]
    \centering
    \includegraphics[width=0.7\textwidth]{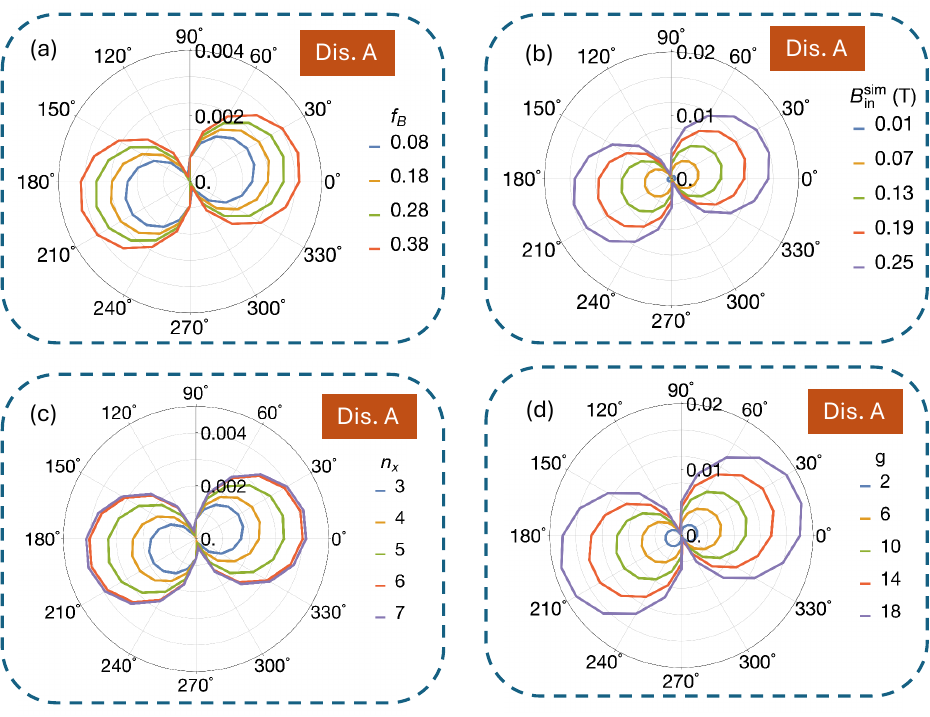}
    \caption{(a)--(d) First-lobe asymmetry $A_1$ as a function of the in-plane-field orientation $\theta$ for representative values of the flux-focusing factor $f_B$,  in-plane field strength $B_{\rm in}^{\rm sim}$, $n_x$,  and the $g$-factor. The same disorder realization, labeled as Dis.~A, and disorder strength, $W_{\rm dis}/t = 1.95$, are used for all panels. In each set of calculations, we vary one parameter at a time while fixing all the remaining parameters: $n_x=4$, $B_{\rm in}^{\rm sim}=0.04~{\rm T}$, $f_B=0.18$, and $g=2$. 
}
    \label{fig:asym_largeWdis}
\end{figure}

Figure~\ref{fig:asym_largeWdis} shows the angular dependence of the asymmetry $A_1$ as $f_B$, $B_{\rm in}^{\rm sim}$, $n_x$, and the $g$ factor are varied. Figures~\ref{fig:asym_largeWdis}(a)--(c) focus on parameters that control the effective dipole flux. Increasing $f_B$, $B_{\rm in}^{\rm sim}$, or $n_x$ monotonically enhances $A_1$, while its principal axis remains nearly aligned with the $0^\circ$--$180^\circ$ direction. These parameters therefore predominantly control the magnitude of the asymmetry, with little effect on its orientation.
 
Within the parameter ranges considered, the additional dipole flux $\Delta\phi_z$ remains much smaller than the background perpendicular flux $\phi_z$ at the relevant side lobes. The common dependence on $f_B$, $B_{\rm in}^{\rm sim}$, and $n_x$, together with the nearly fixed principal axis, supports flux-focusing-induced dipole flux as the primary origin of the angular side-lobe asymmetry.
This behavior characterizes the weak relative-flux regime, $|\Delta\phi_z|\ll|\phi_z|$. At a larger in-plane field, for example, $B_{\rm in}^{\rm sim}=3.0~\mathrm{T}$, corresponding to $B_{\rm in}^{\rm eff}=42~\mathrm{mT}$, $\Delta\phi_z$ becomes comparable in magnitude to $\phi_z$. Consequently, the angular profile shown in Fig.~\ref{fig:Fig2}(e) of the main text, calculated using the same disorder strength and realization, becomes distorted and begins to deviate from the characteristic dumbbell shape.

The dependence on the $g$ factor is shown in Fig.~\ref{fig:asym_largeWdis}(d) for $g\in \{2,6,10,14,18 \}$. For this disorder realization, the magnitude of $A_1$ increases monotonically with $g$, while its principal axis remains nearly aligned with the $0^\circ$--$180^\circ$ direction. Unlike $f_B$, $B_{\rm in}^{\rm sim}$, and $n_x$, however, the $g$ factor does not directly control the additional dipole flux. Instead, it modifies the microscopic current response through the Zeeman coupling. Although Fig.~\ref{fig:asym_largeWdis}(d) shows a qualitatively similar trend for this particular data set, the $g$-factor dependence varies substantially in the weaker-disorder regime and among different disorder realizations, as discussed in Sec.~\ref{sec:asym_results_weaker}.   

Overall, the additional numerical results in Fig.~\ref{fig:asym_largeWdis} support the flux-focusing-induced dipole mechanism underlying the observed angular asymmetry. The systematic dependence on $f_B$, $B_{\rm in}^{\rm sim}$, and $n_x$, together with the nearly fixed principal-axis orientation, is consistent with our physical interpretation and supports the parameter choices adopted in the main analysis. The less universal $g$-factor dependence further motivates the use of a weak Zeeman coupling, for which the calculated angular response most closely reproduces the experimental behavior.

\subsection{Numerical results  for weaker disorder strength}
\label{sec:asym_results_weaker}

In the main text and the preceding section, we use $W_{\rm dis }/t=1.95$. Here, we consider the weaker value $W_{\rm dis }/t=1.1 $, motivated by the estimate in Table~\ref{tab:QuanOscil-para}, to verify that the results are not artifacts of an unusually strong disorder potential. At this disorder strength, the second side lobe remains well resolved, allowing us to analyze both $A_1$ and $A_2$.

We further compare three disorder realizations to distinguish robust qualitative trends from realization-dependent features. In addition to the $ 30 \times 120$ lattice considered above, we examine a larger $ 45 \times 180$ lattice with two additional disorder realizations as a consistency check.

\begin{figure}[ht]
    \centering
    \includegraphics[width=0.95\textwidth]{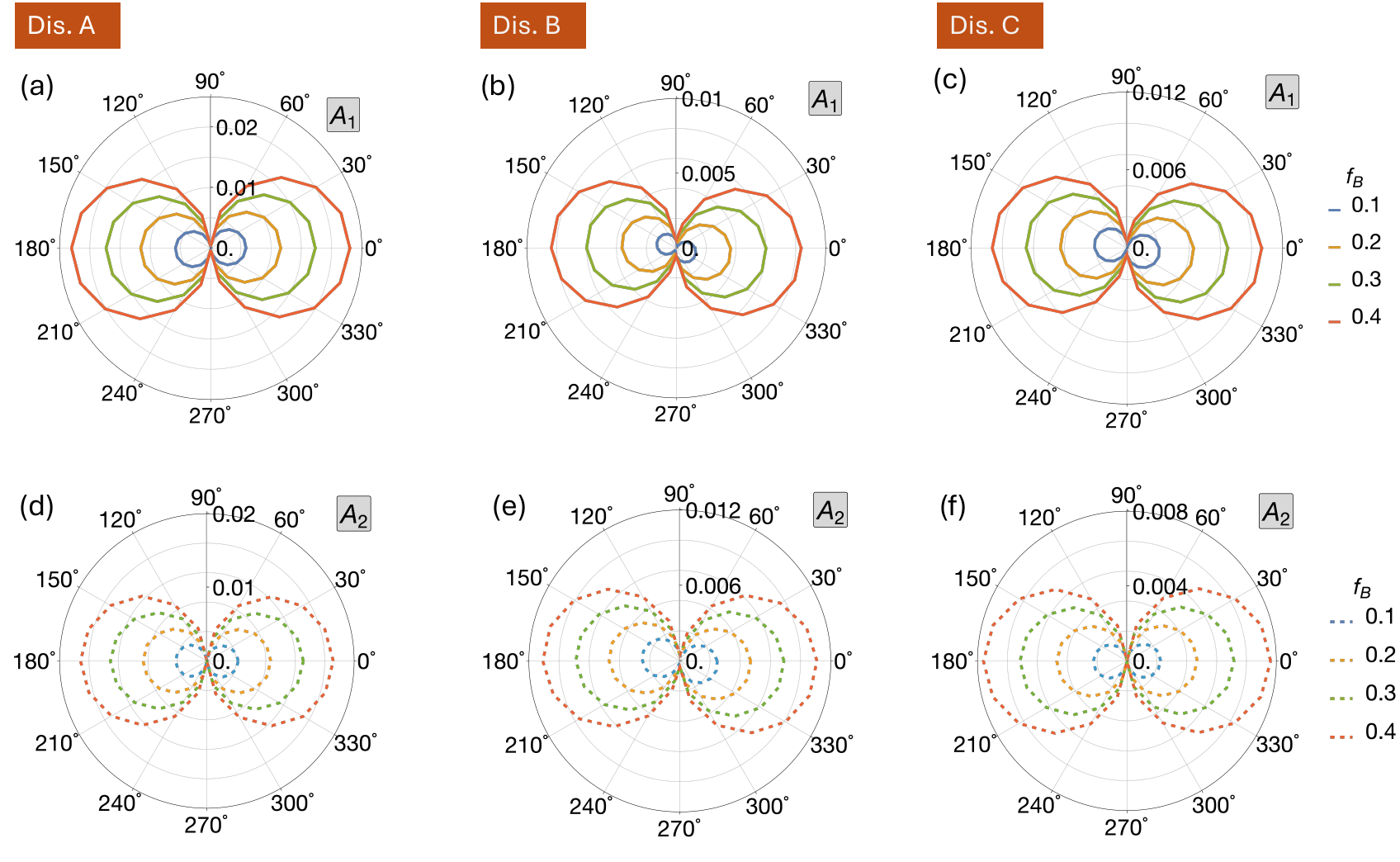}
    \caption{(a,b,c) First- and (e,f,g) second-lobe asymmetry,  $A_1$ and $A_2$, as a function of in-plane field orientation $\theta$, using representative values of flux-focusing factor $f_B$.  The three columns (left to right) correspond to three disorder realizations (labeled as Dis. A, Dis. B, and Dis. C). The adopted remaining parameters  for all panels are fixed as $B_{\rm in}^{\rm sim} = 0.04$~T, $n_x = 5,$ and $g = 2$.
    }
    \label{fig:asym_varfB}
\end{figure}

\begin{figure}[h]
    \centering
    \includegraphics[width=0.95\textwidth]{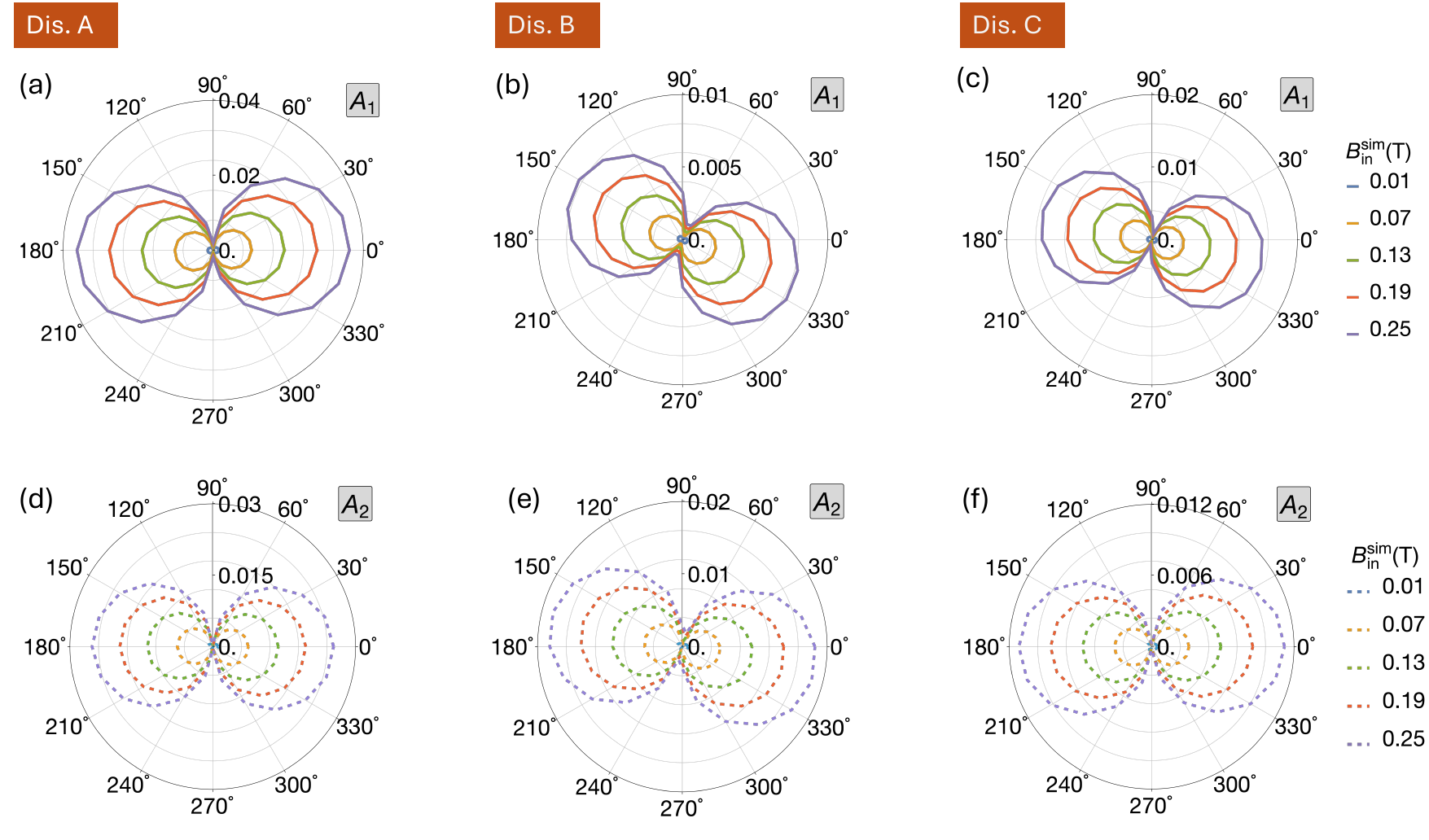}
    \caption{Angular dependence of $A_1$ and $A_2$ for different in-plane field strength $B_{\rm in}^{\rm sim}$. The remaining adopted parameters here are $f_B = 0.1$, $n_x = 5,$ and $g = 2$.
    }
    \label{fig:asym_varBp_small}
\end{figure}

\begin{figure}[h]
    \centering
    \includegraphics[width=0.95\textwidth]{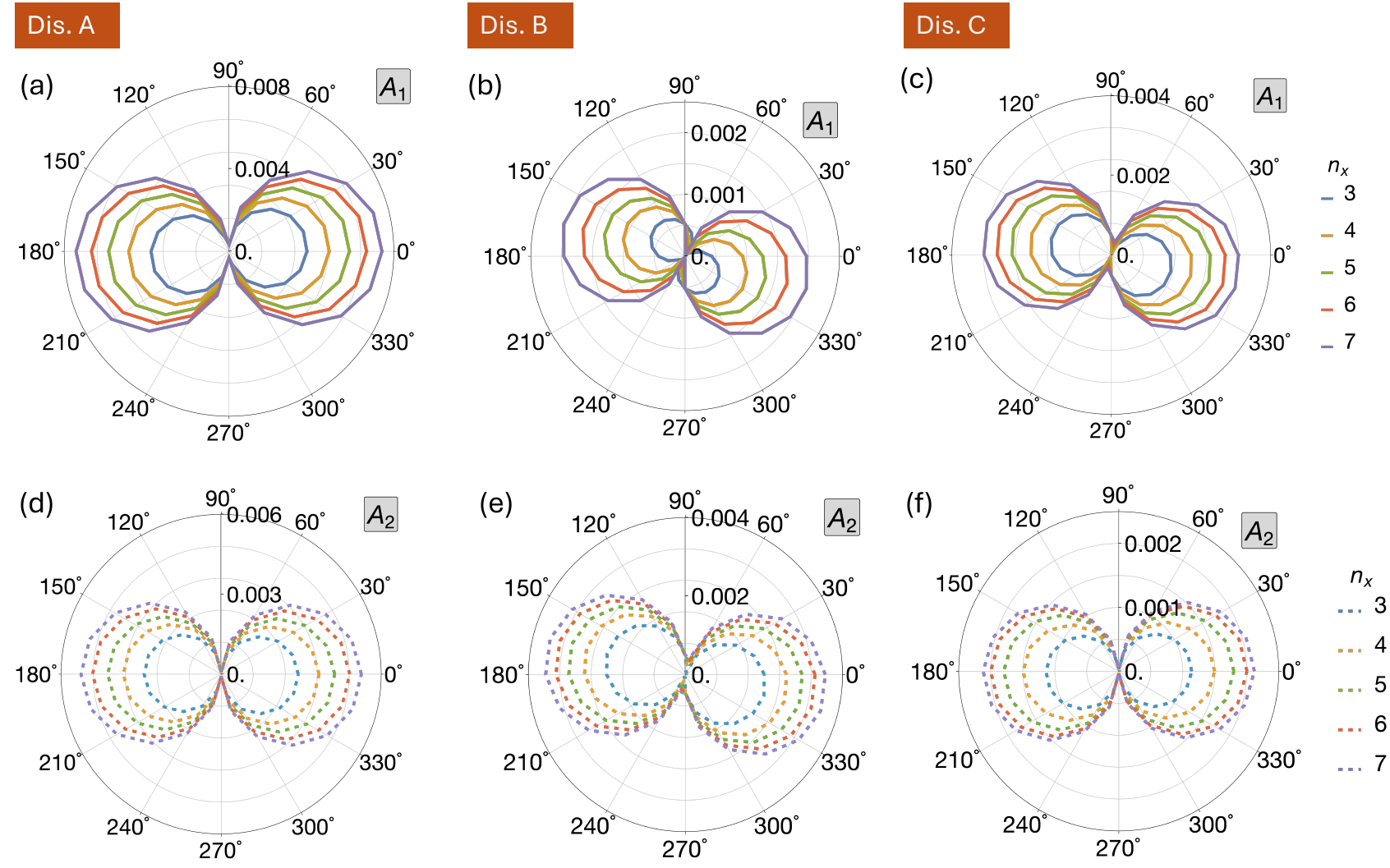}
    \caption{Similar plots to Fig.~\ref{fig:asym_varfB}, but here we vary $n_x$. Among all the calculations, we fix $B_{\rm in}^{\rm sim} = 0.04$~T, $f_B = 0.1$,  and $g = 2$.}
    \label{fig:asym_varnx}
\end{figure}

\begin{figure}[h]
    \centering
    \includegraphics[width=0.95\textwidth]{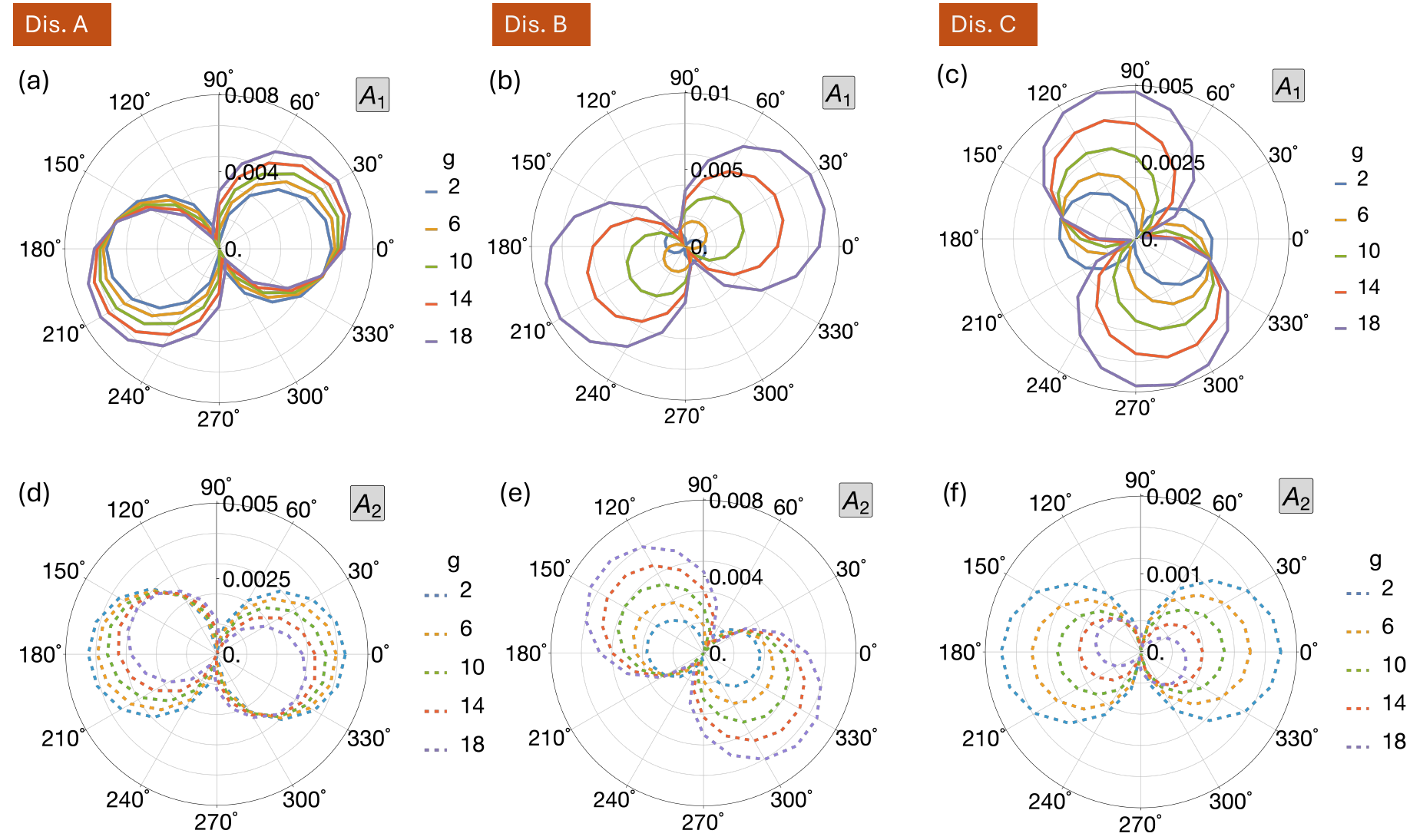}
    \caption{Similar plots to Fig.~\ref{fig:asym_varfB}, except that the $g$-factor is varied. We fix $B_{\rm in}^{\rm sim} = 0.04$ T, $f_B = 0.1$,  and $n_x = 5$ through all the calculations.}
    \label{fig:asym_vargfac}
\end{figure}

\subsubsection{Dependence on disorder realizations, flux dipole, and Zeeman coupling}

Figures~\ref{fig:asym_varfB}--\ref{fig:asym_vargfac} show $A_1$ and $A_2$ for an $N_x\times N_y=30\times120$ normal region and three disorder realizations, denoted by Dis.~A, Dis.~B, and Dis.~C. Dis.~A is the same realization used in the strong-disorder analysis in Sec.~\ref{sec:asym_results}.

The trends in  Figs.~\ref{fig:asym_varfB}--\ref{fig:asym_varnx} remain consistent with those obtained in the strong-disorder regime discussed in Sec.~\ref{sec:asym_results}. 
For all three disorder realizations, increasing $f_B$, $B_{\rm in}^{\rm sim}$, or $n_x$ enhances both $A_1$ and $A_2$, while their principal axes remain nearly aligned with the $0^\circ$--$180^\circ$ direction. These results indicate that parameters controlling the effective dipole flux primarily determine the asymmetry magnitude, while leaving its principal axis nearly unchanged in the relatively weaker-flux regime.

The absolute asymmetry magnitudes, however, depend on the specific disorder realization. A comparison of Dis.~A at the two disorder strengths---results of $W_{\rm dis}/t=1.95$ shown in Fig.~\ref{fig:asym_largeWdis} with the corresponding $W_{\rm dis}/t=1.1$ results in panel (a) of each of Figs.~\ref{fig:asym_varfB}--\ref{fig:asym_vargfac}---further shows that the magnitudes $A_1$ are also sensitive to the overall disorder strength.

Among Figs.~\ref{fig:asym_varfB}--\ref{fig:asym_vargfac} , the $g$-factor dependence is more complex. As shown in Fig.~\ref{fig:asym_vargfac}, the magnitude of $A_1$ increases monotonically with $g$ for all three disorder realizations. The same trend is observed for $A_2$ in Dis.~B, whereas $A_2$ decreases with increasing $g$ in Dis.~A and Dis.~C.
Furthermore, although each angular profile retains a two-fold dependence and its principal axis evolves systematically with $g$, the detailed behavior varies among disorder realizations. In particular, the principal axis can deviate substantially from the $0^\circ$--$180^\circ$ direction. Both the magnitude and orientation of the asymmetry can therefore be modified through the interplay of Zeeman coupling, spin-orbit coupling, and disorder. This comparison distinguishes the systematic flux-dipole-driven behavior from realization-dependent changes in the microscopic current response. The closer agreement with experiment at small $g$, again, motivates the weak Zeeman coupling adopted in the main analysis.

\begin{figure}
    \centering
    \includegraphics[width=0.95\textwidth]{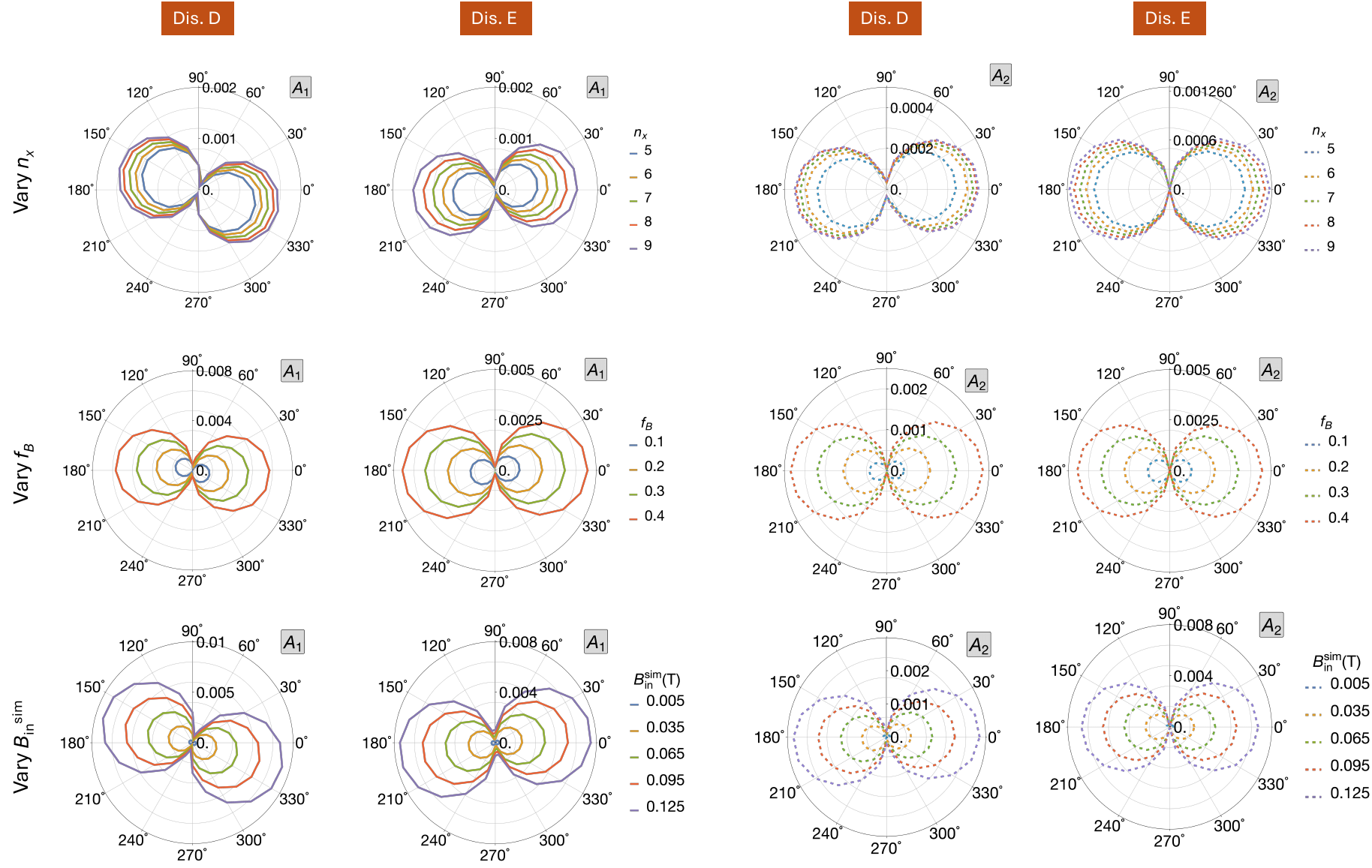}
    \caption{ Angular dependence of the side-lobe asymmetries $A_1$ and $A_2$ computed for the $45\times 180$ lattice with disorder realizations Dis.~D and Dis.~E. Columns 1 and 3 (2 and 4) correspond to Dis.~D (Dis.~E). The top, middle, and bottom rows show the results for varying $n_x$, $f_B$, and $B_{\rm in}^{\rm sim}$, respectively. In this set of calculations, only one parameter is varied at a time: we fix $f_B = 0.1$ and $B_{\rm in}^{\rm sim} = 0.02~{\rm T}$ when varying $n_x$, fix $n_x = 7$ and $B_{\rm in}^{\rm sim} = 0.02~{\rm T}$ when varying $f_B$, and fix $n_x = 7$ and $f_B = 0.1$ when varying $B_{\rm in}^{\rm sim}$. 
    }
    \label{fig:asym_largeN}
\end{figure}

\subsubsection{Numerical results for a larger lattice}

In the analysis above, we primarily considered a junction with $N_x\times N_y=30\times120$. As a consistency check, we consider a larger normal region with $N_x\times N_y=45\times180$ with  $W_{\rm dis}/t=1.1$, using two additional disorder realizations, Dis.~D and Dis.~E. For each realization, we vary $n_x$, $f_B$, and $B_{\rm in}^{\rm sim}$ and analyze the angular dependence of the side-lobe asymmetries. The first, second, and third rows of Fig.~\ref{fig:asym_largeN} show the corresponding results, respectively. The angular profiles of $A_1$ and $A_2$ closely resemble those obtained for the $30\times120$ lattice, confirming that the main qualitative trends remain consistent for a different lattice size.

\begin{figure}
    \centering
    \includegraphics[width=0.95\textwidth]{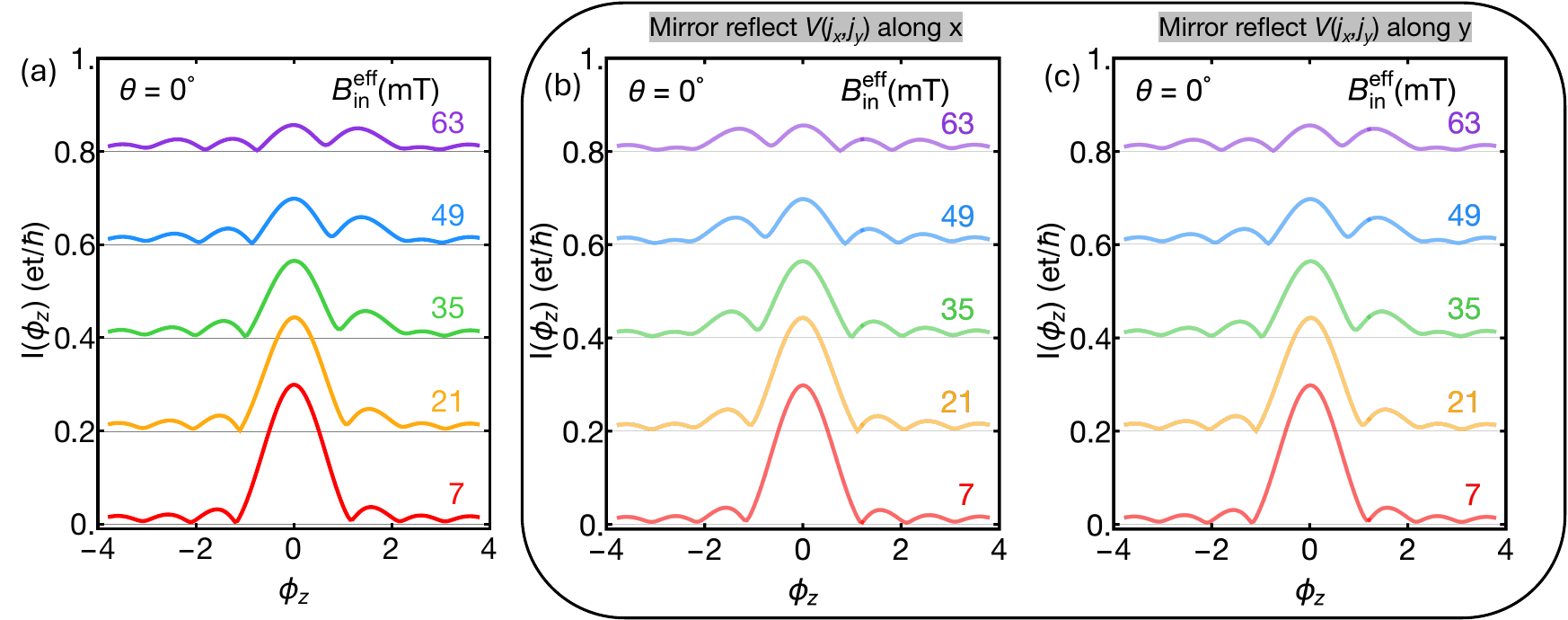}
    \caption{Interference patterns for reflected disorder configurations. (a) Original disorder realization Dis.~A. (b,c) Patterns obtained after reflecting Dis.~A along the $x$ and $y$ directions, respectively, according to Eq.~\eqref{Eq:landscape-reflected}. In all panels, the effective in-plane field $B_{\rm in}^{\rm eff}$ is labeled according to Eq.~\eqref{Eq:B_in-conversion}, and the field orientation is fixed at $\theta=0^\circ$.
}
    \label{fig:Frhfr_flipV}
\end{figure}

\subsubsection{Reflected disorder configurations}

To clarify the role of the spatial disorder profile, we artificially reflect the onsite potential in either the $x$ or $y$  direction:
\begin{subequations}
    \label{Eq:landscape-reflected}
\begin{align}
V(j_x,j_y)&\rightarrow V(N_x-j_x+1,j_y),\\
V(j_x,j_y)&\rightarrow V(j_x,N_y-j_y+1),
\end{align}
\end{subequations}
respectively, and recalculate the interference patterns. The results are shown in Fig.~\ref{fig:Frhfr_flipV}.

Figure~\ref{fig:Frhfr_flipV}(a) shows the interference pattern for the original, unreflected disorder profile. Reflecting this profile in the longitudinal direction  reverses the pattern with respect to $B_z$ [Fig.~\ref{fig:Frhfr_flipV}(b)]. By contrast, transverse reflection leaves the pattern nearly unchanged [Fig.~\ref{fig:Frhfr_flipV}(c)].
This distinction demonstrates that the asymmetry depends not only on the disorder strength but also on the spatial arrangement of the scattering potential relative to the flux dipole.
 
Under longitudinal reflection, the supercurrent encounters the disorder landscape in the reverse order along the transport direction. Because the dipole-induced out-of-plane fields have opposite signs near the two NS interfaces, this transformation interchanges how the scattering landscape couples to the two sides of the flux dipole, thereby reversing the $B_z$ asymmetry. Its effect is therefore analogous to rotating the in-plane field from $\theta=0^\circ$ to $\theta=180^\circ$, which reverses the dipole flux. Transverse reflection, however, preserves the longitudinal ordering of the disorder landscape and only mirrors the current distribution across the junction width. The corresponding phase accumulation is therefore essentially unchanged, and the interference pattern is not reversed.

\begin{figure}
    \centering
    \includegraphics[width=0.95\textwidth]{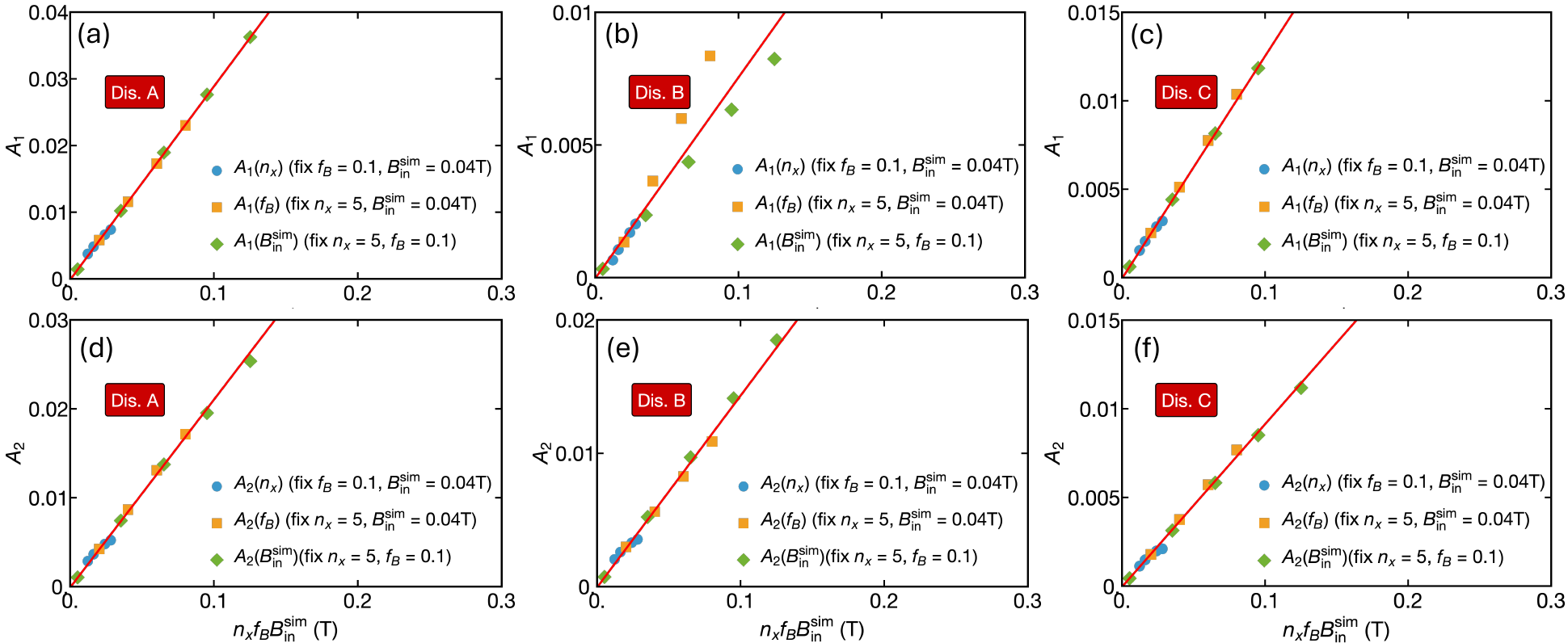}
    \caption{Asymmetries $A_1$ [panels (a)--(c)] and $A_2$ [panels (d)--(f)] for the $30\times 120$ lattice and $W_{\rm dis}/t = 1.1$ exhibit scaling versus $n_x f_B B_{\rm in}^{\rm sim}$. Red lines indicate linear fits to the data. The linear scaling of $A_1$ and $A_2$ is universal across all three disorder realizations. The asymmetry data used in this figure are taken from those in Figs.~\ref{fig:asym_varfB}--\ref{fig:asym_varnx}. 
     }
    \label{fig:scaling}
\end{figure}

\subsection{Scaling with flux dipole and conversion between $B_{\rm in}^{\rm sim}$ and $
B_{\rm in}^{\rm eff}$ 
}
\label{sec:A_scaling_conv_Bin}

\begin{figure}[t]
    \centering
    \includegraphics[width=0.62\textwidth]{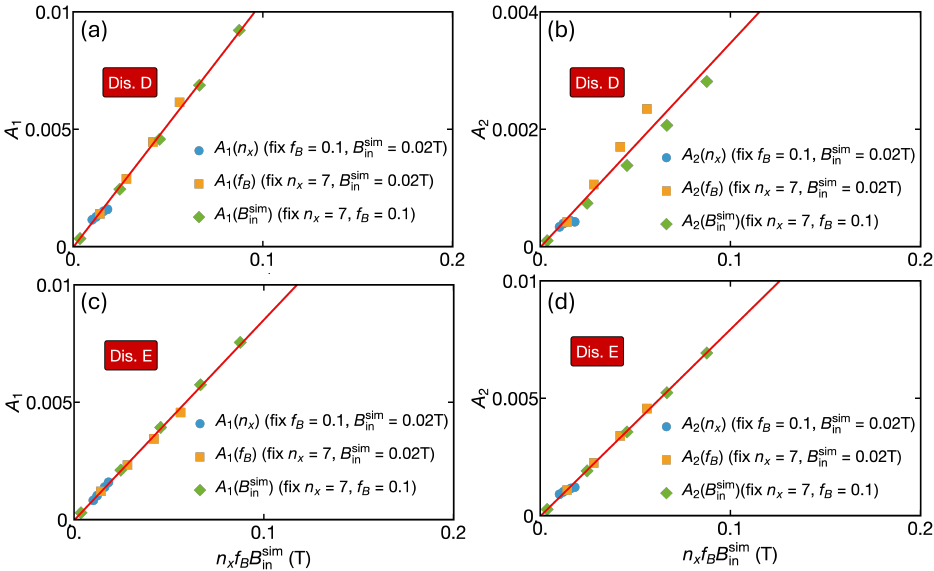}
    \caption{Scaling of $A_1$ [panels (a,c)] and $A_2$ [panels (b,d)] for a larger system, $N_x \times N_y = 45 \times 180$. Two disorder realizations are considered: Dis. D [panels (a,b)] and Dis. E [panels (c,d)]. The analysis is analogous to that in Fig.~\ref{fig:scaling}.    }
    \label{fig:scaling_largeN}
\end{figure}

The similar evolution of $A_1$ and $A_2$ with $f_B$, $B_{\rm in}^{\rm sim}$, and $n_x$ for both the $30\times120$ lattice (Figs.~\ref{fig:asym_varfB}--\ref{fig:asym_varnx}) and the $45\times180$ lattice (Fig.~\ref{fig:asym_largeN}) indicates that the asymmetry is governed primarily by the product of these parameters
\begin{align}
\Delta\phi_z \propto \frac{n_x f_B B_{\rm in}^{\rm sim} \cos\theta}{\Phi_0},
\label{eq:dphi_z}
\end{align}
where $\Delta\phi_z$ denotes the additional out-of-plane flux generated by in-plane-field flux focusing. To test this scaling, in Figs.~\ref{fig:scaling} and \ref{fig:scaling_largeN}, we replot the asymmetry data obtained by varying $B_{\rm in}^{\rm sim}$, $f_B$, and $n_x$, with the remaining parameters fixed, as functions of $n_x f_B B_{\rm in}^{\rm sim}$. 
For each disorder realization and lattice size, the data collapse onto a common linear relation, demonstrating that the asymmetry scales linearly with $n_x f_B B_{\rm in}^{\rm sim} $. This scaling persists across the disorder realizations and lattice sizes considered, providing quantitative support for the flux-focusing mechanism underlying the asymmetry. However, the slope of the linear relation depends on the specific disorder realization.
 
The scaling above also motivates the conversion between the simulated in-plane field $B_{\rm in}^{\rm sim}$, used for the reduced junction size, and the corresponding effective field $B_{\rm in}^{\rm eff}$ in the experimental device. For a smaller dipole area $\Delta A$, a larger field is required to generate the same additional flux $\Delta\phi_z$ and hence a comparable asymmetry. For $\theta = 0 ^\circ$, Eq.~\eqref{eq:dphi_z} can be expressed as
\begin{align}
\Delta\phi_z
\propto
\frac{\Delta A f_B B_{\rm in}^{\rm sim} }{\Phi_0}
=
\frac{\eta_A  A^{\rm sim} f_B B_{\rm in}^{\rm sim} }{\Phi_0}
=
\frac{\eta_A A f_B B_{\rm in}^{\rm eff} }{\Phi_0},
\end{align}
where $A^{\rm sim}$ is the area of the N region and $\eta_A\equiv\Delta A/A^{\rm sim}$ is the fraction occupied by the flux-dipole region in the simulation. We define $B_{\rm in}^{\rm eff}$ as the field that produces the same $\Delta\phi_z$ in the experimental device as $B_{\rm in}^{\rm sim}$ does in the simulated junction, assuming the same $\eta_A$ and $f_B$ in the simulation and experiment.
This gives
\begin{align}
B_{\rm in}^{\rm eff}
=
\frac{A^{\rm sim}}{A}
B_{\rm in}^{\rm sim}.
\label{Eq:B_in-conversion}
\end{align}
Using $A=0.9\times 2.9~\mu\mathrm{m}^2$ and $A^{\rm sim}= 90 \times 360~\mathrm{nm}^2$, the simulated fields of $B_{\rm in}^{\rm sim}=0.5$--$4.5~\mathrm{T}$ correspond to effective fields of $B_{\rm in}^{\rm eff}\approx 7$--$ 63~\mathrm{mT}$. These effective values are used to label the numerical results in the main text and are comparable to the experimental range, $B_{\rm in}=5$--$70~\mathrm{mT}$.


\begin{figure}
    \centering
    \includegraphics[width=0.95\textwidth]{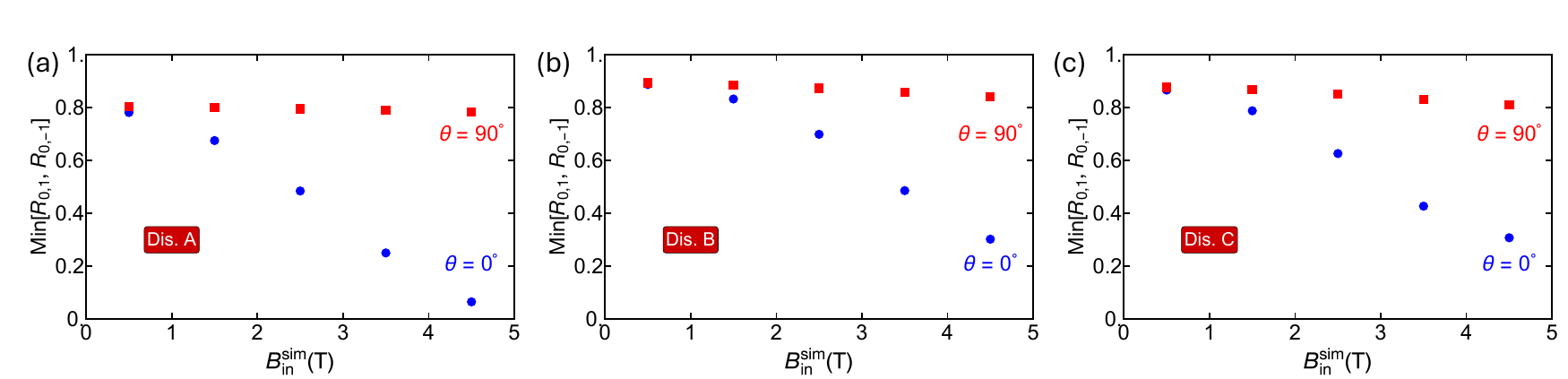}
    \caption{Lobe ratio $\mathrm{Min}\left[R_{0,1}, R_{0,-1}\right]$  plotted as a function of $B_{\rm in}^{\rm sim}$ for $\theta = 0^\circ$ (blue dots) and $\theta = 90^\circ$ (red dots) for three different disorder realizations. $R_{0,1}$ and $R_{0,-1}$ are defined in Eq.~\eqref{eq:loberatio}. 
    }
    \label{fig:MinR}
\end{figure}

\subsection{Fraunhofer-to-SQUID crossover}
\label{sec:crossover}

The distinct evolution of the interference patterns for $\theta=0^\circ$ and $\theta=90^\circ$ is well reproduced by the simulations shown in Fig.~\ref{fig:Fig3} of the main text. To quantify the contrasting suppression of the central lobe for these two orientations, we define
\begin{align}
R_{0,n}\equiv\frac{I^{(0)}-I^{(n)}}{I^{(0)}+I^{(n)}},
\label{eq:loberatio}
\end{align}
where $I^{(0)}$ and $I^{(n)}$ denote the peak currents of the central and $n$th side lobes, respectively. The quantity $R_{0,n}$ measures their normalized peak-height difference. Because an asymmetric interference pattern generally gives $R_{0,n}\neq R_{0,-n}$, we characterize the Fraunhofer-to-SQUID crossover using $\min\left[R_{0,n},R_{0,-n}\right]$. A smaller value indicates that the central lobe is closer in height to at least one of the corresponding side lobes and therefore signifies a more SQUID-like interference pattern.

Focusing on the first side lobes, $n=\pm1$ in Eq.~\eqref{eq:loberatio}, we evaluate $\min \left[R_{0,1},R_{0,-1}\right]$ for $\theta=0^\circ$ and $\theta=90^\circ$ for each of the three disorder realizations. The results are shown in Fig.~\ref{fig:MinR}. For $\theta=0^\circ$, increasing $B_{\rm in}^{\rm sim}$ strongly suppresses $\min \left[R_{0,1},R_{0,-1}\right]$, indicating that the central lobe approaches the height of a first side lobe and that the interference pattern evolves from a Fraunhofer profile toward a SQUID-like profile. By contrast, for $\theta=90^\circ$, $\min \left[R_{0,1},R_{0,-1}\right]$ is only weakly reduced. This orientation-dependent behavior persists across all three disorder realizations. Together with the simulated interference patterns in Figs.~\ref{fig:Fig3}(c) and \ref{fig:Fig3}(d) of the main text, these results reproduce the contrasting evolution  observed experimentally in Figs.~\ref{fig:Fig3}(a) and \ref{fig:Fig3}(b) of the main text.

\FloatBarrier

\section{Phenomenological description of the interference  asymmetry }

Motivated by the numerical analysis of the interference pattern asymmetry in Sec.~\ref{Sec:microscopic}, in particular the  scaling behavior in Sec.~\ref{sec:A_scaling_conv_Bin}, we develop a phenomenological description based on the Dynes--Fulton formalism~\cite{DynesFulton:1971} to   understand our results.  
Based on Ref.~\cite{DynesFulton:1971}, we write the critical current as
\begin{align}
    I(B_z,B_{\mathrm{in}},\theta)	& =\left|\mathcal{I}(B_z,B_{\mathrm{in}},\theta)\right|,\nn
\mathcal{I}(B_z,B_{\mathrm{in}},\theta)	&  =\int_{0}^{L}dx\int_{-W/2}^{W/2}dy\,j(x,y)e^{ik_{B}(x)y},
\label{eq:generalized_DF_disorder}
\end{align}
with  the complex critical current amplitude, $\mathcal{I}$. To account for the spatial inhomogeneity induced by onsite disorder, we introduce the spatially dependent quantity $j(x,y)$ in Eq.~\eqref{eq:generalized_DF_disorder}. Its integral along the transport direction defines the effective critical-current density, $J(y)=\int_0^L dx\,j(x,y)$. In Eq.~\eqref{eq:generalized_DF_disorder}, $k_B(x)$ denotes the local phase gradient, which is proportional to the total perpendicular
magnetic field at position $x$.

\begin{figure}[h]
    \centering
    \includegraphics[width=0.9\textwidth]{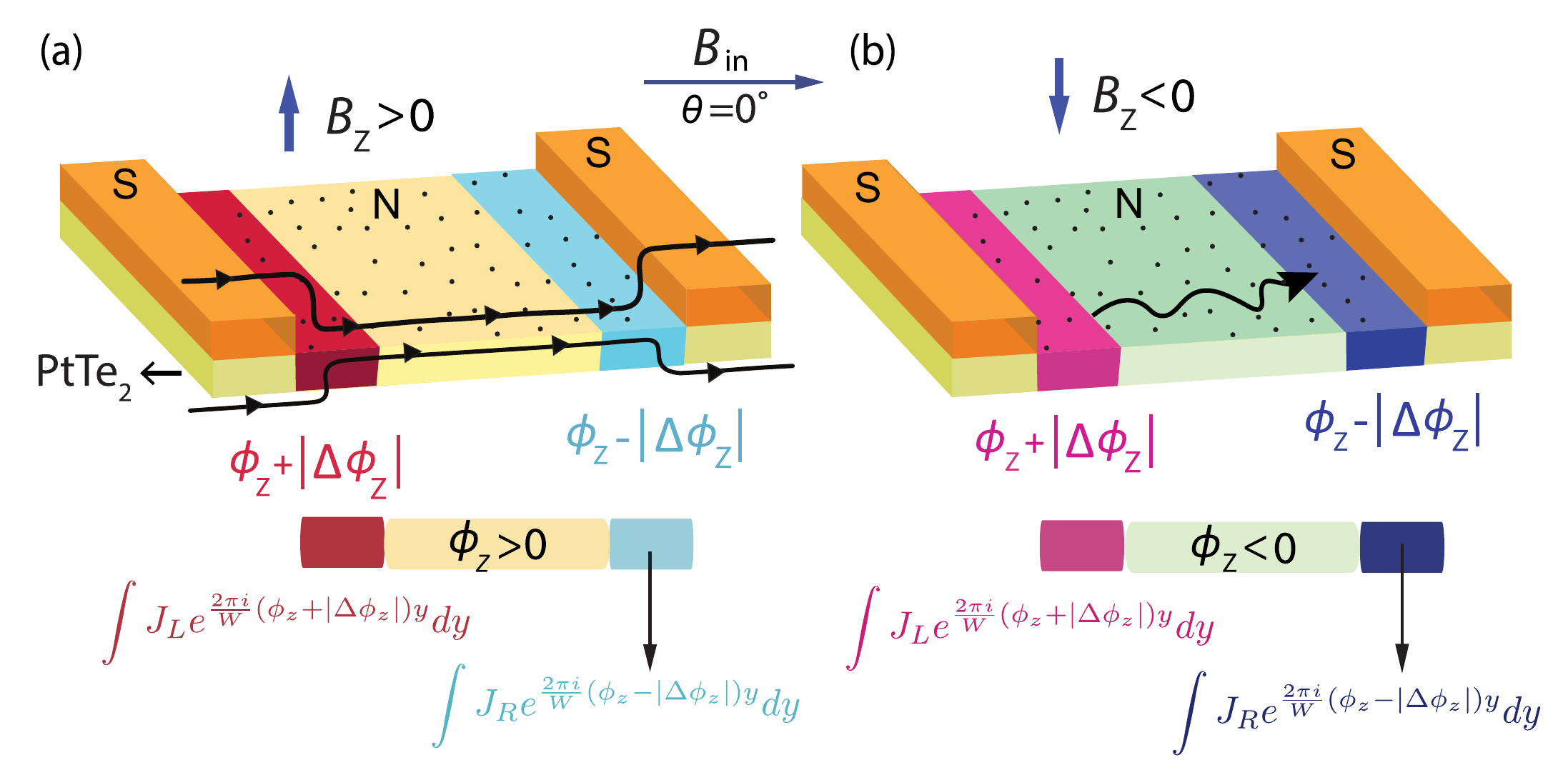}
    \caption{Schematic illustration of the imbalance between the supercurrent contributions from the two interfaces for (a) $B_z>0$ and (b) $B_z<0$. Different colors distinguish the N region subject to different magnetic field fluxes, while the corresponding supercurrent contributions are indicated by the mathematical formula of the same color. 
    }
    \label{fig:current_imbalance}
\end{figure}

To account for the flux dipole, the junction region is separated into left-interface, middle, and right-interface regions, as described in Eq.~\eqref{eq:full_B_dipole} and illustrated in Fig.~\ref{fig:current_imbalance}. 
The corresponding phase gradients are defined as $k_{B}+\Delta k_{B}$, $k_{B}$, and $k_{B}-\Delta k_{B}$, respectively, with $\Delta k_{B}=2\pi\delta B_z n_{x}/\Phi_{0}$ and $\delta B_z = f_B B_{\rm in}\cos(\theta)$ denoting the magnitude of the additional perpendicular magnetic field arising from flux-focusing. 
We denote the effective current distributions obtained by integrating $j(x,y)$ over $x$ within the three regions as $J_L$, $J_M$, and $J_R$, respectively. The complex critical-current amplitude $\mathcal{I}$ then takes the form of 
 \begin{align}
     \mathcal{I}(B_z,B_{\mathrm{in}},\theta)=\int_{-W/2}^{W/2}dy\,e^{ik_{B}y}\left[J_{L}(y,\varphi_L)e^{i\Delta k_{B}y}+J_{M}(y,\varphi_M)+J_{R}(y,\varphi_R)e^{-i\Delta k_{B}y}\right],
\label{eq:three_region_DF}
 \end{align}
where $\varphi_{L,M,R}$ can be determined by the continuity of the supercurrent between different regions.  Equation~\eqref{eq:three_region_DF} can be viewed as a superposition of three  contributions associated with the left-interface, middle, and right-interface regions, weighted by the flux-dipole-induced oscillation factors. 

In the absence of disorder, one gets $J_{L} = J_{R}$ and  the critical current obtained from Eq.~\eqref{eq:three_region_DF} is therefore even under $B_z\rightarrow-B_z$, yielding a symmetric interference pattern. When disorder is included, however, $J_{L}$ and $ J_{R}$ near the two NS interfaces are generally unequal. The combination of this disorder-induced imbalance and the flux dipole generates a component in $\mathcal{I}$ that is odd under $B_z\rightarrow-B_z$.
Therefore, the interference asymmetry arises from their combination:  disorder produces an imbalance between the two interface contributions, whereas the flux dipole converts this imbalance into an odd magnetic-field response.   

This description also provides a consistent explanation for the linear scaling behavior of the asymmetry presented in Figs.~\ref{fig:scaling} and \ref{fig:scaling_largeN} for the parameter regime of small
$B_{\rm in}$, $n_x$, and $f_B$. 
In the present description, this corresponds to   $ |W \Delta k_{B}| \ll 1$, where 
we can expand Eq.~\eqref{eq:three_region_DF} to first order in $W\Delta k_{B}$. 
Away from the nodes of the interference pattern, the even component of $\mathcal I$ remains dominant, while the odd component can be treated as a perturbative correction.
In this regime, the $n$th side-lobe asymmetry of Eq.~\eqref{eq:asymm} is approximately linear in the magnitude of the
additional flux   $\Delta\phi_z$:
\begin{align}
    A_n(B_{\rm in},\theta) = \mathcal{C} |\Delta \phi_z|.
    \label{eq:asymmetry_flux_scaling_short}
\end{align}
Remarkably, Eq.~\eqref{eq:asymmetry_flux_scaling_short} captures the linear scaling $A_{1/2}\propto n_x f_B B_{\rm in}$ across different disorder realizations, as shown in Figs.~\ref{fig:scaling} and \ref{fig:scaling_largeN}. It also explains why the side-lobe asymmetry is maximal for an in-plane field nearly parallel to the current direction, $\theta=0^\circ$, and minimal for a nearly perpendicular field, $\theta=90^\circ$. To leading order, the dimensionless coefficient $\mathcal C$ is independent of $\delta B_z$, but depends on the disorder configuration through $J_L$, $J_R$, and the underlying $B_z$-even interference profile. This dependence qualitatively accounts for the different slopes in Figs.~\ref{fig:scaling} and \ref{fig:scaling_largeN} for different disorder realizations.

The above linear expansion is expected to break down near the Fraunhofer nodes, where the even component of $\mathcal I$ becomes small. It may also become invalid at larger $B_{\rm in}$ or stronger spin-orbit coupling, where modifications of the band structure and current distribution induced by the combined effects of Zeeman splitting and spin-orbit interaction can produce nonlinear responses.

\let\temp\addcontentsline 
\renewcommand{\addcontentsline}[3]{}

\putbib[Ref.bib]
\let\addcontentsline\temp
\end{bibunit}

\end{document}